\documentclass[12pt,preprint]{aastex}

\slugcomment{}

\shorttitle{NLTE solar-type model}
\shortauthors{Short and Hauschildt}

\begin{document}


\title{A NLTE line blanketed model of a solar type star}


\author{C.I. Short}
\affil{Department of Astronomy \& Physics and Institute for Computational Astrophysics, Saint Mary's University,
    Halifax, NS, Canada, B3H 3C3}
\email{ishort@ap.smu.ca}

\author{P.H. Hauschildt}
\affil{Hamburger Sternwarte, Gojenbergsweg 112, 21029 Hamburg, Germany}
\email{phauschildt@hs.uni-hamburg.de}


\begin{abstract}

  We present LTE and NLTE atmospheric models of a star with solar parameters, 
and study the effect of treating many thousands of Iron group lines out of LTE 
on the computed atmospheric structure, overall absolute flux distribution, and 
the moderately high resolution spectrum in the visible and near UV bands.  Our 
NLTE modeling includes the first two or three ionization stages of 20 chemical 
elements, up to and including much of the \ion{Fe}{0}-group, and includes about 
20000 \ion{Fe}{1} and \ion{}{2} lines.  We investigate separately the effects 
of treating the light metals and the \ion{Fe}{0}-group elements in NLTE.  Our 
main conclusions are that 1) NLTE line blanketed models with direct multi-level 
NLTE for many actual transitions gives qualitatively similar results as the 
more approximate treatment of \citet{anderson_89} for both the \ion{Fe}{0} 
statistical equilibrium and the atmospheric $T_{\rm kin}$ structure, 2) models 
with many \ion{Fe}{0} lines in NLTE have a $T_{\rm kin}$ structure that agrees 
more closely with {\it LTE} semi-empirical models based on center-to-limb 
variation and a wide variety of spectra lines, whereas LTE models agree more 
with semi-empirical models based only on an LTE calculation of the \ion{Fe}{1} 
excitation equilibrium, 3) the NLTE effects of \ion{Fe}{0}-group elements on 
the model structure and $F_\lambda$ distribution are much more important than 
the NLTE effects of all the light metals combined, and serve to substantially 
increases the violet and near UV $F_\lambda$ level as a result of NLTE Fe 
over-ionization.  These results suggest that there may {\it still} be important 
UV opacity missing from the models.  However, the choice of the species and 
multiplet dependent van der Waals broadening enhancement also plays a 
significant role in determining whether LTE or NLTE models provide a close fit 
to the near UV flux level.  We also find that the RMS deviation of the shape 
of the rectified high resolution synthetic spectrum from that of the the 
observed spectrum is not significantly affected by the inclusion of NLTE 
effects.

\end{abstract}


\section{Introduction}


Of the many fronts on which models of stellar atmospheres and spectrum formation 
need to be made more realistic,
one is the treatment of the thermodynamic state of the gas and the radiation field.
A particular difficulty in this regard is that complex atoms and ions such as those of \ion{Fe}{0} 
and the \ion{Fe}{0}-group elements have a particularly rich term structure 
with many transitions that need to be accounted for in a realistic equilibrium solution.
The very thing that makes these species problematic also gives impetus to their
accurate treatment: their many transitions provide a dense line opacity that
veils the spectrum and partially controls the value of the broad-band emergent flux,
especially in the violet and UV bands of late-type stars.  
We emphasize at the outset that the accuracy of atmospheric models depends on 
both the physical realism of the modeling, and on the quality and completeness
of the input physical data.  The inadequacies of the latter are documented
in, for example, \citet{kurucz02}.  Here we study the effect of improvement in the former while
holding the latter fixed.  The goal of this paper is to 
ask and answer this question: what are the effects on the model of a late-type star and its 
computed spectrum of treating the equilibrium state of the \ion{Fe}{0}-group elements and many of their 
transitions more realistically than has been done in the past, while holding
fixed the other degrees of realism that typify current models of late-type stars.  We 
have partially addressed this question for the \ion{Fe}{0} equilibrium in red giant
stars (\citep{short_h03}).  Here we address the question more completely
for \ion{Fe}{0} and other \ion{Fe}{0}-group elements in a late-type main sequence (MS)
star, namely one that has the parameters of the Sun.
 
\paragraph{}

Determination of the Fe abundance in the closest late-type star,
the Sun, has been particularly problematic 
(see \citet{kostik_sr96} for a very thorough discussion).  
Related to this has been the problem of correctly modeling the overall flux 
level in the near UV band (3000 to 4000\AA)
\citep{kurucz92a}, which is veiled by thousands of weak \ion{Fe}{1} lines.
The most widely used theoretical atmospheric models for late-type stars are 
the ATLAS series of models \citep{kurucz94a}
and the MARCS (most recently NMARCS) series of models
\citep{plez_bn92}, both of which make use of the simplifying assumption
of Local Thermodynamic Equilibrium (LTE).
As a result of the vast expansion of the input line lists in the early 1990's,
these theoretical LTE models have been more successful at fitting the solar flux distribution in the
near UV band than have semi-empirical models such as that of 
Holweger \& Mueller \citep{holmul} (see, for example, \citet{blackwell_ls95}).
However, the details of line formation not only affect the predicted 
flux level directly by way of its effect on opacity, but also indirectly
by way of the effect of opacity on the radiative equilibrium structure of the 
atmosphere.  
In the case of the \ion{Fe}{1} spectrum and the UV flux level, it
has been found that the approximation of LTE ``conspires'' to off-set errors in 
the atmospheric structure, and thus
provide a deceptively accurate prediction of the UV flux level (see, for 
example, \citet{kostik_sr96} and references therein for a thorough discussion).
\citet{anderson_89} calculated theoretical NLTE models of the solar atmosphere with
many opacity sources treated in NLTE and found that NLTE departures in the Fe 
equilibrium significantly affect the atmospheric T structure.  The latter work is
a significant development, however, it treats the NLTE radiative transfer problem
is a more approximate way ((the {\it multifrequency/multigray algorithm}) rather than
the more direct way employed here.

\paragraph{}

The atmospheres of late-type MS stars are translucent over large 
enough path lengths that non-locally determined 
radiative rates often dominate collisional rates for many transitions, causing departures 
from local thermodynamic
equilibrium (LTE).  Previous NLTE investigations of \ion{Fe}{1} with
atomic models of more limited scope than that used here 
(see \citet{shchukina_t01} for a recent example) 
have found that the the weak \ion{Fe}{1} lines become significantly weaker
due to NLTE over-ionization.
Therefore, {\it both} the atmospheric structure and the emergent spectrum
should be calculated in the more general statistical equilibrium (SE), in
which a set of coupled equations is solved for the rate at which every
energy level of every ionization stage of every species is populated
and de-populated by various collisional and radiative processes.

\paragraph{}

   As a result of computational constraints, previous investigations of NLTE effects in the atmospheres
of the Sun and other late-type stars have treated at most a few hundred 
spectral lines in SE while treating most of the strongest lines and most of
the ``haze'' of weaker lines as LTE ``background'' opacity.  The notable exception is
the modeling of \citet{anderson_89} described above.  However, for complex species
such as \ion{Fe}{1} and \ion{}{2} the method of \citet{anderson_89}
solves the SE equations for model states composed of many real states
(the {\it multifrequency/multigray algorithm}) rather than for real states. 
Recently, \citet{short_hb99} have modified the multi-purpose atmospheric modeling and spectrum
synthesis code, {\tt PHOENIX} \citep{hauschildt_b99}, to greatly increase the number of chemical species that
are treated in NLTE SE.  As a result, over 100\,000 spectral lines throughout the spectrum, including all of the 
strongest lines and many of the weaker lines that blanket the UV band, 
are now treatable in self-consistent NLTE.  This includes $\sim$6900 lines of 
\ion{Fe}{1}, $\sim$13600 lines of \ion{Fe}{2}, and $\sim$35000 lines due to the first two
ionization stages of the \ion{Fe}{0}-group elements Ti, Mn, Co, and Ni.  Even this is a small fraction of
the millions of spectral lines that collectively control the emergent flux, but it is a
significant step forward in improving the realism of the models.  The purpose of this study is to
calculate theoretical models and synthetic spectra for the Sun, {\it considered as a star}, 
with the newly expanded NLTE treatment to assess the affect
that large scale NLTE line blanketing has on the theoretical model structure 
and the synthetic
spectrum in the problematic violet and near UV bands.  In Section \ref{sec_model} we describe the computational 
modeling;
in Section \ref{sec_reslt} we present our results, and we re-iterate our main 
conclusions in Section \ref{sec_con}.

\section{Modeling \label{sec_model}}

{\tt PHOENIX} makes use of a fast and accurate Operator Splitting/Accelerated Lambda Iteration
(OS/ALI) scheme to solve self-consistently the radiative
transfer equation and the NLTE statistical equilibrium (SE) rate equations for many species and overlapping transitions
\citep{hauschildt_b99}.
Recently \citet{short_hb99} have greatly
increased the number of species and ionization stages treated in
SE by {\tt PHOENIX} so that at least the lowest two stages of 24 elements,
including the lowest six ionization stages of the 20 most
important elements, including Fe and four other \ion{Fe}{0}-group elements, are now treated in NLTE.  
\citet{short_hb99} contains details
of the sources of atomic data and the formulae for various atomic processes.
Table 1 shows which species have been treated in NLTE in the modeling presented here, and how many $E$ levels and $b-b$ (bound-bound) transitions are included in 
SE for each species.  For the species treated in NLTE, only levels connected by
transitions of $\log gf$ value
greater than -3 (designated primary transitions) are included directly in the SE rate equations.  
All other transitions of that species (designated secondary transitions) are calculated
with occupation numbers set equal to the Boltzmann distribution value with excitation 
temperature equal to the local kinetic temperature, multiplied by the ground state 
NLTE departure co-efficient for the next higher ionization stage. 
We have only included in the NLTE treatment those ionization stages that are non-negligibly 
populated at some depth in the Sun's atmosphere.  As a result, we only include the
first one or two ionization stages for most elements.  

\paragraph{}
NLTE effects can depend sensitively on the adopted values of atomic parameters that affect the 
rate of collisional and radiative processes.  Atomic data for the energy levels and b-b transitions have been taken from \citet{kurucz94b} and \citet{kurucz_b95}.
We have used the resonance-averaged Opacity Project \citep{seaton_ymp94} data of
\citet{bautista_rp98} for the ground-state photo-ionization cross sections of \ion{Li}{1}-\ion{}{2}, \ion{C}{1}-\ion{}{4}, \ion{N}{1}-\ion{}{6}, \ion{O}{1}-\ion{}{6}, \ion{Ne}{1}, \ion{Na}{1}-\ion{}{6}, \ion{Al}{1}-\ion{}{6}, \ion{Si}{1}-\ion{}{6}, \ion{S}{1}-\ion{}{6}, \ion{Ca}{1}-\ion{}{7}, and \ion{Fe}{1}-\ion{}{6}.  
For the ground states of all stages of P and Ti
and for the excited states of all species, we have used the cross sectional data previously incorporated into
{\sc PHOENIX}, which are from \citet{reilman_m79} or those compiled by \citet{mathisen84}.  We account for
coupling among {\it all} bound levels by electronic collisions using cross sections calculated with the 
formula of \citet{allen73}. 
We do not use the formula of \citet{vanreg62} for pairs of levels that are connected by a permitted
radiative transition because we have found that doing so leads to rates for
transitions within one species that are very discrepant with each other,
and this leads to spurious results.  
The cross sections of ionizing collisions with electrons are calculated with the formula of \citet{drawin61}.  
We describe in this paper a perturbation analysis of collisional 
rates.

\paragraph{}

   Table 2 shows the three levels of realism with which we model the equilibrium 
state of the gas and the radiation field. 
Unless otherwise
noted, the realism of a synthetic spectrum calculation is always consistent with
that of the input model used.  
Our NLTE modeling includes two levels
of realism: 1) NLTE treatment for H, He, and important 
light metals up to, but {\it not} including, the \ion{Fe}{0}-group elements (designated NLTE$_{\rm Light}$ models), 
and 2) the same as the NLTE$_{\rm Light}$ models except that the \ion{Fe}{0}-group elements Ti, Mn, Fe Co, and Ni 
are also included in the NLTE treatment (designated NLTE$_{\rm Fe}$ models). 
The \ion{Fe}{0}-group elements play a special role
in the atmospheres and spectra of late-type stars \citep{thevenin_i99}; because of their spectacularly rich term
structure a neutral or low ionization stage \ion{Fe}{0}-group element contributes between 
one and two orders of magnitude more lines to the spectrum than the corresponding
stage of any lighter element.  Finally, we note that all of the models in
Table 2 also include many tens of millions additional lines from many atoms,
ions and diatomic molecules in the approximation of LTE.  
Note that H$^-$, which is an important source of continuous opacity in late-type MS stars, is treated
in LTE here.

\subsection{Modeling limitations \label{modlim}}

\paragraph{Atomic parameters:} 

We do not make any attempt here to fine-tune the atomic parameters
that control the formation of individual spectral lines.
Such fine tuning of oscillator strengths ($gf$ values) and damping constants
($\gamma$ values) is necessary for the derivation of
accurate abundances of particular species from the fit to particular 
spectral lines.  Our purpose is to investigate the collective consequence 
of including NLTE effects in the formation of many spectral lines.  
We presume that errors in the atomic data in the line list of \citet{kurucz92a}
are random and will not bias the apparent collective effect of massive-scale
NLTE.

\paragraph{Model structure:} 

We note that late-type stars generally have 
a chromospheric temperature inversion in their outer 
atmosphere.
 Because our models are in radiative/convective
 equilibrium, they have monotonically decreasing $T_{\rm kin}(\tau)$ structures.
Therefore, the cores of strong lines that are sufficiently opaque to
remain optically thick very high in the atmosphere, and for which the
emergent radiation field is dominated by local thermal conditions where
the line forms, will not be
accurately reproducible with our model.  We expect to predict too little
flux in the cores of such lines.  Also, as $\lambda$ decreases the
atmosphere becomes increasingly opaque such that, in the case of the Sun,  
the entire pseudo-continuum at $\lambda < 2500$\AA~ 
forms {\it above} the depth of minimum temperature 
($T_{\rm min}$).  Therefore, we must expect that theoretical atmospheric
models will necessarily increasingly under-predict the emergent flux there,
and we restrict our spectral fitting to $\lambda > 3000$\AA.  For late-type
stars with particularly active chromospheres, chromospheric pseudo-continuum
emission may contaminate the flux at $\lambda$ values longer than $3000$\AA,
so caution must be used when comparing theoretical and observed flux. 
 
\paragraph{Homogeneity:}
 
   Based mainly on high spatial and temporal resolution studies of the Sun
and on more limited studies of other late-type stars, 
it is expected that late-type stellar atmospheres generally possess a 
broad variety of
structures (starspot umbrae and penumbrae, granules, \ion{Ca}{2} $K$ cells),
 any one of which poses a modeling challenge in its own right.
Progress in our understanding of late-type stellar atmospheres at this level 
requires
the inclusion of physical processes that are beyond the usual ability of 
long time integration, disk-integrated flux spectroscopy to test, 
such as 2D and 3D hydrodynamics and radiative transfer,
and magneto-hydrodynamics (MHD),
and are not included in our modeling.  Furthermore, a one 
dimensional model that is fit to spectral features over a broad range
of wavelengths may not be a meaningful average of the actual structure 
because the relative amount of flux contributed by hotter and cooler components
of the atmosphere depends non-linearly on the temperature the wavelength 
 (see \citet{asplund00} and \citet{shchukina_t01}) for an
example of modeling inhomogeneities).  Here we restrict ourselves to
the traditional limitations of stellar modeling and 
compare computed and observed disk integrated {\it flux}.

\subsection{Model parameters}

Here, we model a late type MS star that has the 
parameters of the Sun for the purpose of studying the differential effects 
of NLTE for a case where the stellar parameters are best known.
In keeping with the assumptions that are still often employed when modeling
stars, our models are homogeneous and
obey radiative/convective equilibrium.  We compute theoretical
structures, taking into account NLTE effects to the degrees shown in
Table 2.  
We adopt the parameters $T_{\rm eff}=5777$ and $\log g=4.4377$, and
the abundances, $[A/H]$, of \citet{grev_ns92}.  Convection carries
most of the flux at the bottom of our model, and we treat it with
the approximation of mixing length theory.  We have adopted a 
value for the mixing length parameter of one pressure scale height 
($l/H_{\rm P}=1.0$).  Our choice of
the thermal micro-turbulent velocity dispersion, $\xi_{\rm T}$, is 1.0 km s$^{-1}$
at all depths, and is larger than the value that 
is sometimes adopted when modeling the solar {\it intensity} spectrum that arises 
from a particular solar feature.  From comparison of intensity and 
disk integrated flux spectra of the Sun, it has been established that 
the adoption of a non-zero $\xi_{\rm T}$ value is at least partially a 
``fudge''
factor that approximately accounts for line broadening that arises from
small scale inhomogeneity.  Our choice of $1.0$ km s$^{-1}$ reflects
the greater range of inhomogeneity that must be so ``fudged'' when modeling 
disk integrated spectra, and is consistent with the value often employed
in models of other solar-type stars.  Finally, we find that 
line profiles in the flux spectrum can be well fitted without recourse to 
macro-turbulent broadening (another parameter suspected to be ``fudge''),
which we, therefore, neglect.

\paragraph{}

Of particular significance for modeling the near UV flux is the Iron 
abundance, $[{\rm Fe}/{\rm H}]$.  The value of $[{\rm Fe}/{\rm H}]$ has
been controversial, with some groups finding a ``low'' value of $~7.51$
(see, for example, \citet{holweger_kb95}) and other groups finding a ``high'' value of 
$~7.63$ (see, for example, \citet{blackwell_ls95}).  The apparent reasons for 
the discrepancy
have been contentious and involve the complexities and uncertainties of 
equivalent width ($W_\lambda$) measurement, atomic data measurement, 
the structure of the model used for line strength calculations, and the
treatment of line formation physics.  For an exhaustive review of the situation,
see \citet{kostik_sr96}.  Here we adopt the value of \citet{grev_ns92}, 7.50,
which is close to the meteoritic value.  It is not the purpose of
the present investigation to determine the solar value of $[{\rm Fe}/{\rm H}]$.
Rather, our goal is to determine how NLTE line formation, including that
of \ion{Fe}{1} and \ion{}{2}, affects the predicted near UV flux level for a ``reasonable''
choice of $[{\rm Fe}/{\rm H}]$.

\section{Results \label{sec_reslt}}

\subsection{Model Structure}

\begin{figure}
\plotone{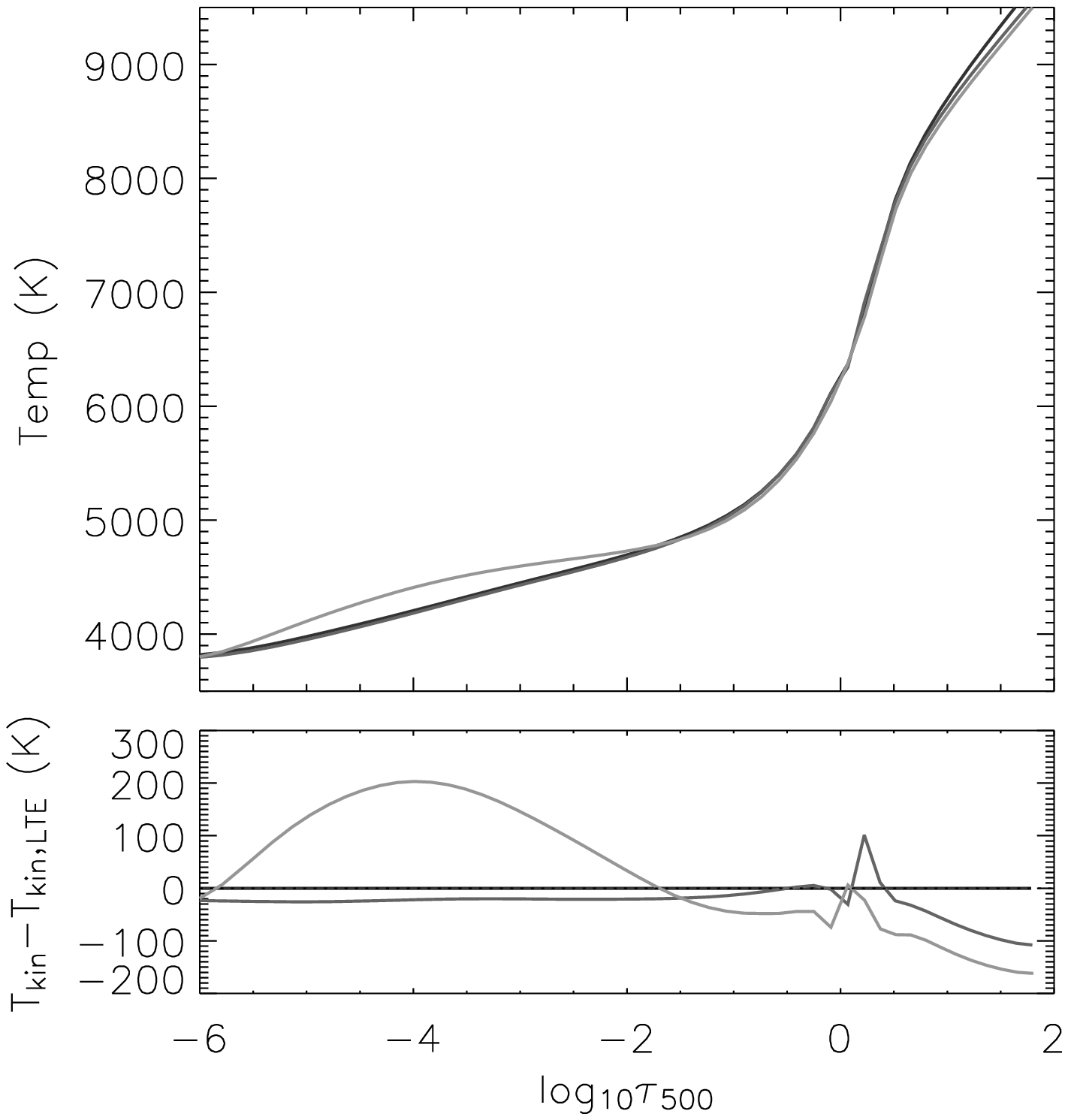}
\caption{Temperature ($T_{\rm kin}$) structure of theoretical atmospheric models of the 
Sun as a function of continuum optical depth
at 500 nm ($\tau_{500}$), computed with: 
LTE (dark line), light metals in NLTE (NLTE$_{\rm Light}$) (medium line),
light metals and \ion{Fe}{0}-group elements in NLTE (NLTE$_{\rm Fe}$) (light line).
Upper panel: Absolute $T_{\rm kin}$, lower panel: $T_{\rm kin}$ relative to
that of the LTE model; dotted line: 0 K.  
\label{sun_mods}}
\end{figure}

Fig. \ref{sun_mods} shows the computed kinetic temperature ($T_{\rm kin}$) structure 
for all of the solar models as a function of the logarithm of the optical depth due to 
continuous opacity at 500 nm ($\log\tau_{500}$).  
We have also plotted the depth-wise
temperature difference between the models and the LTE model.
The triangular kinks in the $T$ differences that are seen in the lower
panel around $\log\tau_{500}\approx 0$ are due to an inflection in the
LTE $T_{\rm kin}$ structure where the NLTE $T_{\rm kin}$ distribution 
crosses the LTE distribution twice within a narrow range of $\log\tau_{500}$. 
Late-type dwarf atmospheres are expected to be close to LTE because of the relatively
large influence of collisions with respect to the radiation field in controlling
the equilibrium state of the gas in a relatively compressed, cool atmosphere. 
We find the NLTE deviations in $T_{\rm kin}$ are everywhere less than 250K. 
The $T_{\rm kin}(\log\tau)$ structure of the NLTE$_{\rm Fe}$ model has a larger and 
more depth dependent deviation from the LTE model than does the NLTE$_{\rm Light}$ model.
It is $\approx$ 150 K cooler than the LTE model at the bottom of the photosphere
($\log\tau\approx 2$), is warmer than the LTE model by as much as 200 K 
in the range $-5 < \log\tau < -2$, and is has steeper surface cooling at
$\log\tau < -5$.  

\paragraph{}

  NLTE effects in line formation have an effect on the equilibrium structure of the atmosphere
by way of the requirement of radiative equilibrium.  The deviation that arises is a result
of the combination of many factors: that some fraction of line photons are scattered rather
than thermally absorbed and emitted, thereby reducing the line's effectiveness as a coolant near
the depth of formation; that the depth at which the line source functions 
become thermalized is altered by the presence of scattering, and that the depth of
formation of lines is altered by changes in line opacity that arise due to NLTE effects on population
number, both of which alter the depth at which a 
line is an effective coolant; and that the intensity at line
center is altered by NLTE changes in the value of the line source function.  
In the case of the NLTE$_{\rm Fe}$ model, the well-known LTE line blanketing effect of 
back-warming is slightly reduced in the $-4 < \log\tau < 2$ range, making for a slightly shallower 
$T_{\rm kin}$ structure.  At the same time, the well-known effect of line cooling near the surface
is enhanced (see \citet{anderson_89} for a detailed discussion of NLTE line formation effects on
the radiative equilibrium of the Sun, and \citet{mihalas_78} for a basic discussion of 
LTE and NLTE effects of lines on theoretical atmospheric structure).

\subsubsection{Comparison with other models \label{s_modelcomp}}

\begin{figure}
\plotone{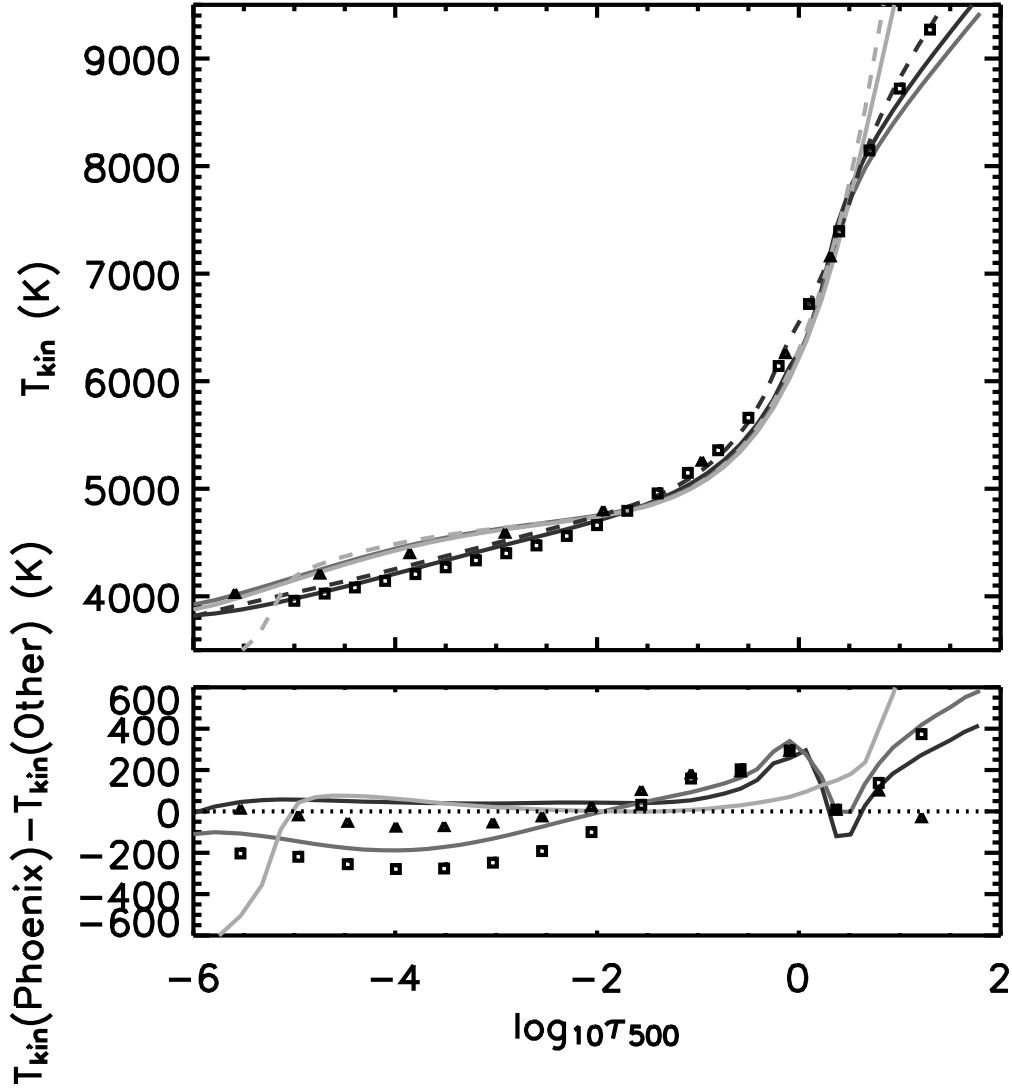}
\caption{Temperature ($T_{\rm kin}$) as a function of continuum optical depth
at 500 nm ($\tau_{500}$) of atmospheric models of the 
Sun from various sources.  Theoretical models of other authors: dashed lines; 
{\sc PHOENIX}
models with parameters corresponding to those of other authors: 
solid lines.
Dark lines: {\sc ASUN} and LTE models with parameters of \citet{kurucz92a}, 
medium lines: 
NLTE$_{\rm Fe}$ models with parameters of \citet{kurucz92a}, light lines: 
{\sc PAM} and NLTE$_{\rm Fe}$ with parameters of \citet{anderson_89}. 
Semi-empirical models of: \citet{holmul} (\sc{HOLMUL}): triangles, and of 
\citet{grev_s99} (\sc{HOLMUL'}): squares. 
Upper panel: Absolute $T_{\rm kin}$, lower panel: Difference between 
$T_{\rm kin}$ of other author and that of corresponding {\sc PHOENIX} model;
dotted line: 0 K. 
\label{sun_compmods}}
\end{figure}

Fig. \ref{sun_compmods} shows the comparison between the $T_{\rm kin}(\log\tau_{\rm 500})$ 
structure of our models and that of four seminal 
models from other investigators: the LTE model of \citet{kurucz92a} computed
with the {\sc ATLAS9} code (designated
{\sc ASUN}), the NLTE line
blanketed model of \citet{anderson_89} computed with the {\sc PAM} code
(designated {\sc PAM}), the LTE 
semi-empirical model of \citet{holmul} (designated {\sc HOLMUL}), and a 
modification of the {\sc HOLMUL} model of \citet{grev_s99} (designated 
{\sc HOLMUL'}).  
To facilitate this comparison we re-calculated our {\sc PHEONIX} models
with the values of $[{{\rm Fe}\over{\rm H}}]$ and ``secondary'' stellar 
parameters ($l/H_{\rm P}$ and $\xi_{\rm T}$) used by the authors of these other models.
Table 3 shows the values of these parameters used for each of the comparisons. 
Differences among the $T_{\rm kin}(\log\tau_{\rm 500})$
structures will reflect not only differences in the equilibrium $T_{\rm kin}(r)$structure 
computed by each set of investigators, but also differences in the computed
value of the continuous opacity at 500 nm at each depth, $\chi_{\rm c, 500}(r)$,and hence in $\tau_{\rm 500}(r)$. 
However, both are relevant to the prediction of the emergent flux, so a comparison
of $T_{\rm kin}(\log\tau_{\rm 500})$ is relevant.

\paragraph{{\sc ASUN}: }

The {\sc ASUN} model is a purely theoretical LTE model that was calculated 
with the {\sc ATLAS9} atmospheric modeling code \citet{kurucz92a}. 
The atomic and molecular data for line blanketing opacity that is input to
our {\sc PHOENIX} calculation is that of {\sc ATLAS9}.  Therefore, the
comparison to {\sc ASUN} is an indicator of differences that
arise due to the methodology and implementation of the solution to the LTE 
atmospheric modeling problem. 
The difference between our LTE model and {\sc ASUN} is less than 
100 K for $\log\tau_{\rm 500}$ values greater than 0.
Near the bottom of the model the difference approaches
400 K.  However, near the bottom of the model the $T_{\rm kin}(\log\tau)$ structure becomes 
steep so that small differences in the slope lead to large $T_{\rm kin}$
differences at a particular depth.  Furthermore, the $T_{\rm kin}$ structure near the bottom of the model is affected
by the treatment of convection, which, in the {\sc PHOENIX} model, carries a significant
fraction of the flux below $\log\tau\approx 0.18$.  We emphasize again that the 
{\sc PHOENIX} models that we are comparing with {\sc ASUN} have been 
computed with the same value of $l/H_{\rm P}$ used to compute the {\sc ASUN} model, $1.25$.

\paragraph{\sc PAM:}

The only previous theoretical model of the solar atmosphere with many 
transitions, including those of \ion{Fe}{0} in NLTE is that of \citet{anderson_89}, 
who used the {\sc PAM} code.  This modeling is based on a much more approximate treatment of the
NLTE rate equations that the approach used here 
(the {\it multifrequency/multigray} algorithm).   
\citet{anderson_89} found that $T$ increased by as much as $\approx 100$ K 
in the $-5 < \log\tau_{500} < -3$ range (his opacity scale) in NLTE, in general
agreement with our NLTE$_{\rm Fe}$ results.  From 
of Fig. \ref{sun_compmods} we find that our NLTE$_{\rm Fe}$ model agrees closely with {\sc PAM}, 
deviating by by less than 100 K in the $-5 < \log\tau_{500} < 0$ range.
This is a very significant result given the large differences
in the way in which {\sc PHOENIX} and {\sc PAM} treat the NLTE SE problem.  Based on a 
careful analysis of each type of transition's contribution to the radiative
equilibrium, \citet{anderson_89} concluded that strong Fe transitions dominate the thermal 
equilibrium in this range.  
\citet{anderson_89} also found, as we do, a sharper
drop off in $T$ in the $-6 < \log\tau_{500} < -5$ range than in LTE, which he
attributed to CO cooling.  However, a comparison to our modeling is difficult
because the models of \citet{anderson_89} extend to $\log\tau_{500} = -8.5$ and we
treat CO in LTE.   

\paragraph{}

\begin{figure}
\plotone{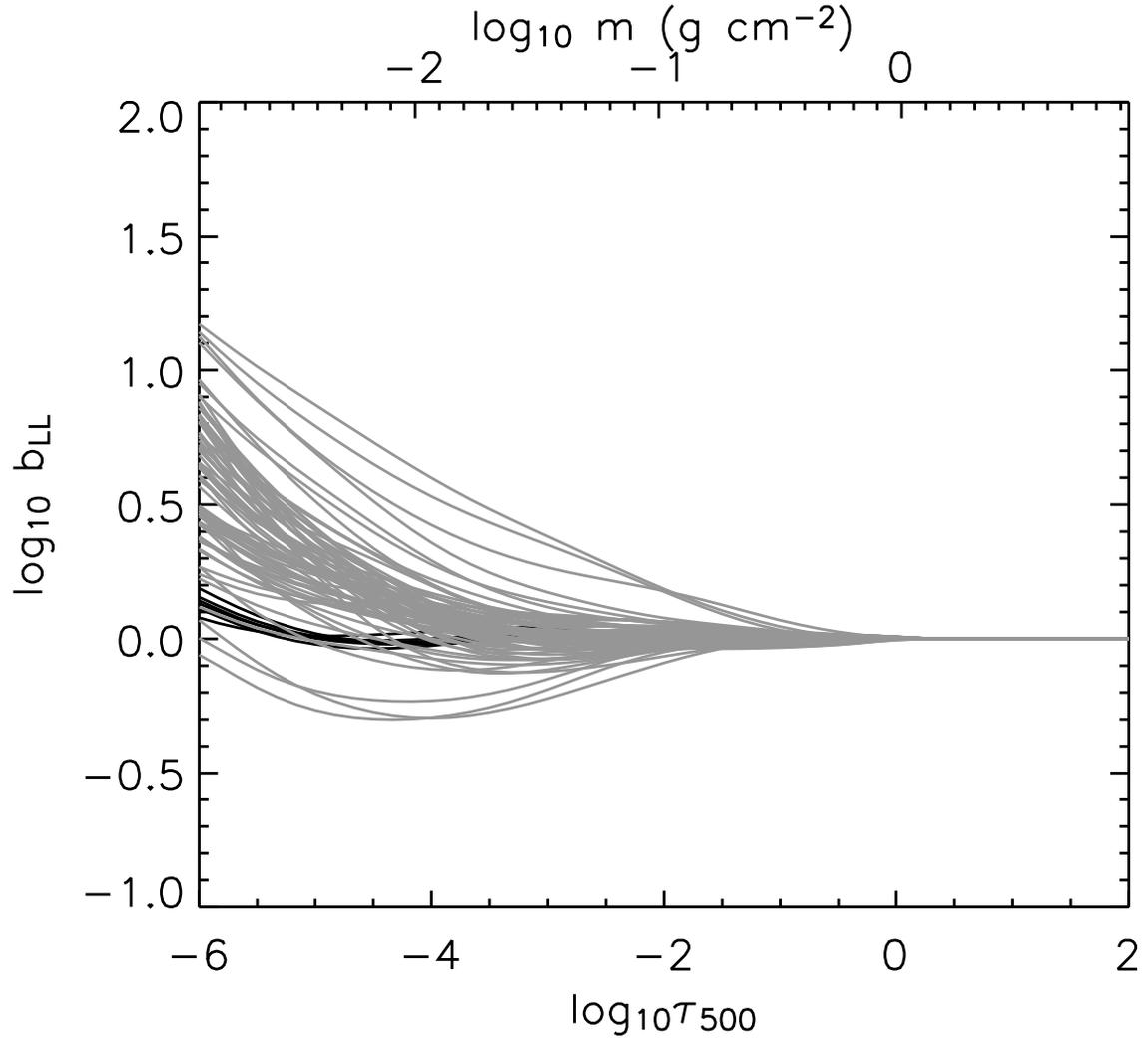}
\caption{Ratio of upper to lower level NLTE departure co-efficients $b_{\rm LL}=b_{\rm u}/b_{\rm l}$
for select transitions of \ion{Fe}{2} that fall within the model states {\it f} (dark lines) and 
{\it g} (light lines) of
\citet{anderson_89}.  Lower $x$-axis: $\log\tau_{\rm 500}$ for consistency with Figs. \ref{sun_mods},
\ref{sun_compmods}, and \ref{sun_bi}.
Upper $x$-axis: logarithm of column mass density for comparison with Fig. 11a of \citet{anderson_89}. 
\label{anders_bll}}
\end{figure}

In Fig. 11a of \citet{anderson_89} the long dashed lines show the ratio,
$b_{\rm LL}$, of the departure coefficient of the upper level, $b_{\rm u}$, to 
that of the lower level, $b_{\rm l}$, for select permitted transitions of 
\ion{Fe}{2},
that take place {\it within} the {\it f} and {\it g} model states of 
Anderson's formalism.  (For transitions that fall with a model state, Anderson
calculates $b_{\rm LL}$ (also denoted in \citet{anderson_89} as $b_{\rm ul}$) values 
using the Equivalent
Two Level Atom (ETLA) formalism; see Eq. 28 and 29 of \citet{anderson_89}).  
The NLTE departure co-efficient of an excitation state, $b_{\rm i}$, is
defined to be the ratio of the actual occupation number of the state,
$n_{\rm i}$ to that that it would have if calculated with an LTE distribution, 
$n_{\rm i}^*$ (the Boltzmann excitation and the Saha ionization distributions 
with excitation and ionization temperatures equal to the local thermal 
temperature).
An important qualification for the definition of $b_{\rm i}$ used both in {\sc PHOENIX}
and in {\sc PAM} is that $n_{\rm i}^*$ is calculated using the {\it actual}
NLTE values of both the electron density ($n_{\rm e}$) and the ground state
occupation number of the next higher ionization stage.  
The significance of the ratio $b_{\rm LL}$ is that it measures the amount
by which the population {\it ratio} of two levels within an ionization stage departs from 
the local
Boltzmann ratio, and is thus insensitive to NLTE departures in the ionization 
balance. 
We have re-created Anderson's quantity $b_{\rm LL}$ for our models by extracting the  
departure co-efficients, $b_{\rm i}$, from the {\sc PHOENIX} output
for excitation states of \ion{Fe}{2}
that would qualify as falling within Anderson's model states {\it f} and {\it g}
(defined by the energy level with respect to the ground state, 
$\chi$, being within the range 4.0 to 5.5 $\times 10^4$ 
cm$^{-1}$, and 6.0 to 8.0 $\times 10^4$ cm$^{-1}$, respectively)
that are connected by permitted transitions,
 and calculating the ratios $b_{\rm LL}=b_{\rm u}/b_{\rm l}$.  These
ratios are shown as a function of depth in Fig. \ref{anders_bll}, which
may be directly compared to Anderson's Fig. 11a.  To facilitate the
comparison, the $x$-axis has been graduated with the logarithm of column mass
density, $\log m$, as well as $\log\tau_{\rm 500}$.  Note that for a gas in LTE
$b_{\rm LL}$ would equal unity everywhere. 
Unfortunately, \ion{Fe}{2} is the only species for which Anderson provides the
mapping of actual transitions to model states (Fig. 2 of \citet{anderson_89}),
so we were unable to perform a similar comparison for \ion{Fe}{1}.  
Nevertheless, our quantities $b_{\rm LL}$ for Anderson's \ion{Fe}{2}
{\it f} and {\it g} model states show the same qualitative behavior as
do Anderson's: under-population of  upper levels with respect to lower levels
for {\it some} transitions
as compared to LTE in the depth range $-2.5 < \log m < -1$ by factors
of as low as 0.4 dex, and over-population for {\it all} transitions at depths of $\log m < -2.5$
by factors approaching 1.0 dex.  For some of our transitions the ratio $b_{\rm LL}$ reaches
higher values than those of \citet{anderson_89}.  However, we note that Anderson's value
reflects the statistical equilibrium computed for one model state that represents the 
collective behavior of all the transitions shown in our Fig. \ref{anders_bll}. 

\paragraph{Semi-empirical models {\sc HOLMUL} and {\sc HOLMUL'}: }

Ideally, theoretical models should match semi-empirical models, although we 
do not expect such a match with our models because of the amount of physics
that has been neglected (see section \ref{modlim}).  Nevertheless, the
importance of the neglected physics can be estimated by assessing the 
quality of match to theoretical models of limited realism. 
The temperature structure of the {\sc HOLMUL} model 
was inferred from fits of {\it LTE} $F_\lambda$ and 
intensity ($I_\lambda$) distributions to spectral line profiles and the
continuum intensity at disk center and near the limb.  
Although {\sc HOLMUL} is an LTE model, we have chosen to compare it to our {\it NLTE$_{\rm Fe}$}
model to study the difference between a semi-empirical model without a chromospheric
$T$ inversion and 
our most realistic theoretical model in radiative-convective equilibrium.
In any case, it can be seen from the upper panel of Fig. \ref{sun_compmods} that {\sc HOLMUL}
is in closer agreement to our NLTE$_{\rm Fe}$ model than to our LTE model throughout
the upper atmosphere. 
Indeed, {\sc HOLMUL} deviates from our NLTE$_{\rm Fe}$ model by less than
100 K in the $-6 < \log\tau_{500} < -1.5$ range.   
Again, the largest differences are near the bottom of the model where 
the $T_{\rm kin}(\log\tau)$ becomes steep and convection plays a role. 
The {\sc HOLMUL'} model is an adjustment
made to {\sc HOLMUL} by \citet{grev_s99} to reconcile solar $[{\rm Fe}/{\rm H}]$
values derived from low and high excitation \ion{Fe}{1} lines.  It is
systematically 200 K cooler than {\sc HOLMUL} throughout the upper atmosphere.
From the upper panel of Fig. \ref{sun_compmods} we note that {\sc HOLMUL'} provides 
a much closer to fit to our LTE $T_{\rm kin}$ structure than to our NLTE$_{\rm Fe}$
structure above a $\log\tau_{\rm 500}$ of -1.5.  

\paragraph{}
It is noteworthy that of the two
semi-empirical models, one closely tracks the NLTE$_{\rm Fe}$ and the other
the LTE theoretical model throughout the upper atmosphere.  We emphasize again that
{\sc HOLMUL'} is based on an LTE calculation of the Fe line strengths, 
whereas {\sc HOLMUL} is based on a wider variety
of diagnostics, including center-to-limb variation.  Therefore, this is possibly
a demonstration of the ``self-fulfilling prophecy'' nature of adopting 
LTE versus NLTE in the treatment of Fe in semi-empirical models as described by 
\citet{rutten86}.

\subsection{Absolute flux distribution \label{sect_absflux}}

\begin{figure}
\plotone{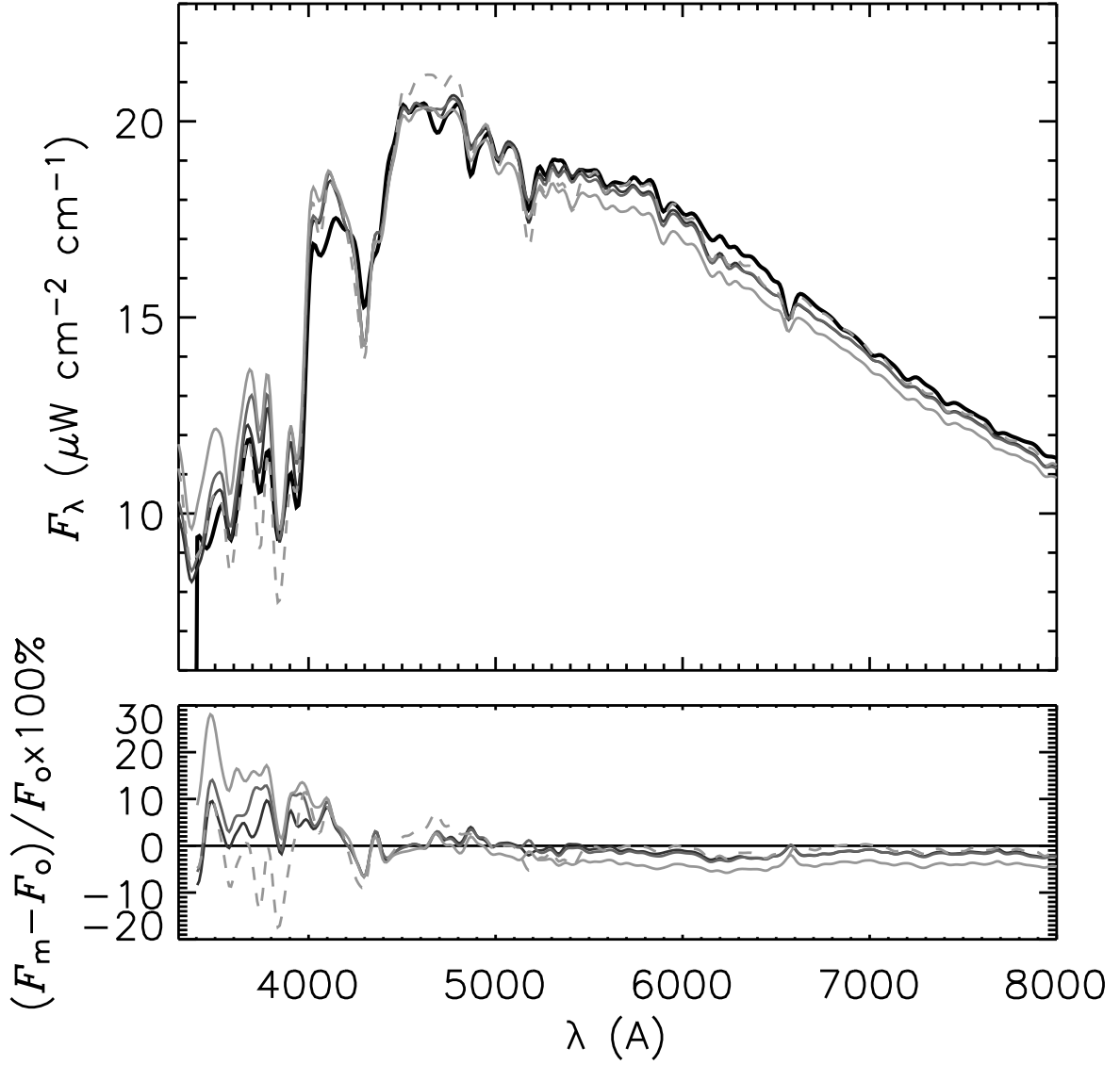}
\caption{Comparison of the absolute global flux distribution ($F_\lambda(\lambda)$) 
of the Sun measured by \citet{necklabs} (thick black line) and synthetic
distributions computed for LTE (dark line), NLTE$_{\rm Light}$ (medium line), and NLTE$_{\rm Fe}$ (light line) models, 
the latter with  a $\gamma_6$ enhancement factor of unity (solid line) and of 1.8 (dashed line).  
Upper panel: Absolute $F_\lambda$, lower panel: the difference between the model ($F_{\rm m}$) and observed ($F_{\rm o}$) $F_\lambda$ distributions, as a
percentage of $F_{\rm o}$.  \label{sun_flxall}} 
\end{figure}

Fig. \ref{sun_flxall} shows the comparison of the observed solar flux distribution,
$F_\lambda(\lambda)$, as measured by \citet{necklabs} and the model flux 
distributions.  We also show the deviation of the computed flux
from the observed flux as a percentage of the observed flux.  
We have re-sampled and convolved the \citet{necklabs} data so that it has a 
uniform $\Delta\lambda$ of 50\AA.  We also convolved our medium 
resolution ($R=200\,000$) synthetic 
$F_\lambda$ distribution with a Gaussian of a FWHM value of 50 \AA~ 
to facilitate the comparison.

\paragraph{}

   The LTE and NLTE models are in close agreement with the
observed $F_\lambda$ distribution on the Rayleigh-Jeans side of the
solar $F_\lambda$ distribution where line blanketing is less severe.  
However, all models become increasingly discrepant with the observed
flux for $\lambda \stackrel{<}{\sim} 5500$ \AA~ where the 
$F_\lambda$ distribution is increasingly affected by line
blanketing.  Both errors in the atomic parameters that
affect line formation, and errors in the physics of line formation, have
an increasingly large effect on the computed $F_\lambda$ distribution
with decreasing $\lambda$ due the increasing line opacity.  While the
former source of error is presumably random, the latter may be systematic.
Indeed, in the $3500 <\lambda < 5000$ \AA~ region the 
more realistic NLTE models provide a fit that is increasingly  
{\it worse} than the LTE models as the level of NLTE realism is increased.
Whereas the LTE model systematically over-predicts $F_\lambda$ by less than
$10\%$ for $\lambda < 4500$\AA, the NLTE$_{\rm Fe}$ model over-predicts
$F_\lambda$ by as much as $30\%$.  This may indicate that the adoption
of LTE masks some other inadequacy in the models.  One possibility
is that, despite the addition of tens of millions of theoretical lines
by \citet{kurucz92a}, the model opacity is still incomplete in the UV band.

\paragraph{}

In this regard 
it is important to note that the treatment of line broadening has a significant
effect on the calculated $F_\lambda$ level on a broad-band scale because of the
collective effect of damped lines on the emergent flux, especially in the heavily
blanketed UV region.  Our value of the $VW$ broadening enhancement parameter, $E_6$,
which is taken to be the same for all lines, has been tuned to
provide a close match to the wing profiles of many damped lines in the solar
spectrum (see Section \ref{sect_hires} and Figs. \ref{sun_atl1} and \ref{sun_atl2}).  
We have found that we get the best
fit to the profiles of damped lines by adopting an enhancement factor of unity.
This very interesting point is elaborated upon in Section \ref{sect_hires}.
  As a result, the collective
opacity of damped lines in our spectrum synthesis is smaller than that of
previous calculations such that our calculated $F_\lambda$ is larger than that
computed by other investigators using the same atmospheric parameters.
For comparison purposes, we have also calculated an 
LTE spectrum with a more traditional enhancement factor of 1.8 and shown it in
Fig. \ref{sun_flxall}.  As expected, the introduction of the enhancement factor
depresses $F_\lambda(\lambda)$ in the UV such that the LTE model {\it under-}predicts
$F_\lambda$ there.  We find that an LTE model of canonical 
solar atmospheric parameters provides a closer fit to $F_\lambda(\lambda)$ with 
an $E_6$ value of unity ({\it ie.} {\it no}
enhancement), which is consistent with our result that the enhancement is not needed
to fit the detailed line profiles.         

\paragraph{Near UV band: }

\begin{figure}
\plotone{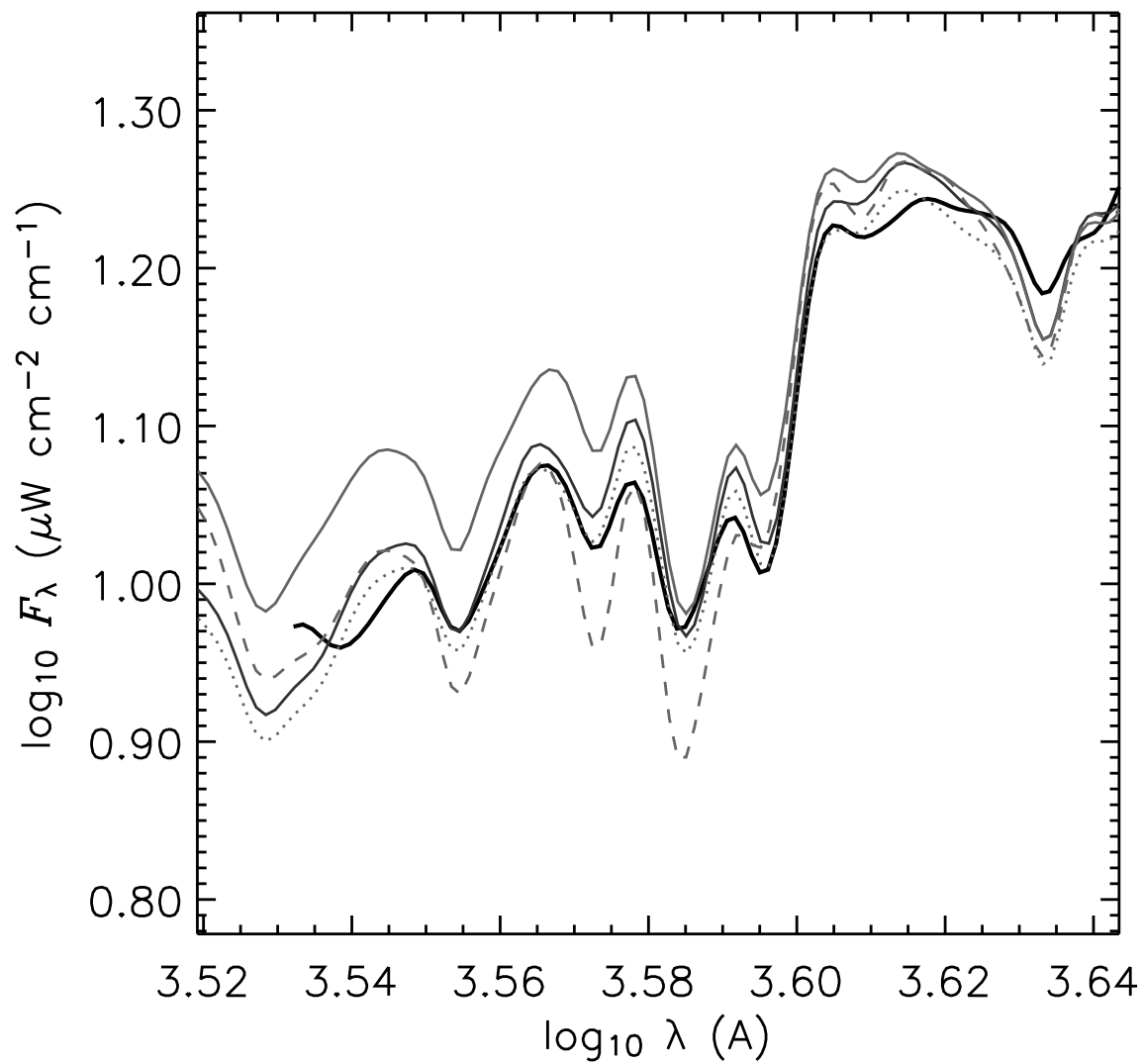}
\caption{Comparison of observed \citet{necklabs} (thick black line) and 
computed violet and UV band logarithmic flux
distributions ($\log F_\lambda(\log\lambda)$).  Theoretical distributions are shown for
LTE (thin dark line), NLTE$_{\rm Light}$ (medium line), and  
NLTE$_{\rm Fe}$ (light line) models.  LTE spectrum synthesis
with NLTE$_{\rm Fe}$ model: dotted line, 
NLTE$_{\rm Fe}$ spectrum synthesis with $\gamma_6$ enhancement factor of 1.8: dashed line. 
\label{sun_flxuv}}
\end{figure}

Fig. \ref{sun_flxuv} shows the comparison of the observed and the computed
$\log F_\lambda(\log\lambda)$ distributions in the $\lambda < 5000$ \AA~ region.
Discrepancies between the computed $F_\lambda$ distributions and the
observed distribution, and among the computed $F_\lambda$ distributions
themselves, are largest in this region.  It should be noted that
traditionally solar models have been too bright in the UV band, which
discrepancy has been described as the ``UV flux problem'' \citep{kurucz90}.
However, \citet{kurucz92a} found that solar models fit the UV band $F_\lambda$
distribution much better when the previous atomic line  
lists, which consisted mainly of lines for which atomic data had been
measured in the laboratory, are supplemented by millions of theoretically
predicted lines from atomic model calculations.  We are using the more
complete line lists of \citet{kurucz92a} in our models.

\paragraph{}

We also computed $F_\lambda$ in LTE with the NLTE$_{\rm Fe}$ model
as input.
Such a calculation is
internally inconsistent, but allows us to assess the relative importance
of direct NLTE line formation effects and indirect NLTE atmospheric
structure effects in accounting for the difference between the 
LTE and NLTE$_{\rm Fe}$ UV $F_\lambda$ distributions.  The resulting $F_\lambda$
distribution is also shown in Fig. \ref{sun_flxuv}, where it can be seen that it
differs negligibly from the self-consistent $F_\lambda$
distribution computed with the LTE model.  This indicates that
it is direct NLTE effects on the line formation through the radiative transfer 
and statistical equilibrium that account for most of the difference between
the $F_\lambda$
distributions of the LTE and NLTE$_{\rm Fe}$ models, rather than the differences
in the atmospheric structure.  Given the small extent of NLTE deviation seen
in Fig. \ref{sun_mods} this is not surprising.
Finally, we also computed LTE and NLTE$_{\rm Fe}$ $F_\lambda(\lambda)$ distributions
using the {\sc HOLMUL'} model for the input atmospheric structure.  Because
{\sc HOLMUL'} is cooler than our LTE model by as much as 
$\sim 100$ K at $\log\tau$ values less than -1, it is expected
to yield a fainter UV band $F_\lambda$ level.  However, we found that
the $F_\lambda(\lambda)$ computed with {\sc HOLMUL'} differed negligibly
from that computed with the {\sc PHOENIX} models, for both the LTE and 
NLTE$_{\rm Fe}$ set-up.  The negligible difference in predicted 
$F_\lambda(\lambda)$ reflects the small
difference in the $T_{\rm kin}$ structures throughout the outer atmosphere
among the models.

\paragraph{}

   The reason for the increased UV flux in the case of the NLTE$_{\rm Fe}$ models can
be seen in Figs. \ref{fe_grot}, \ref{sun_pp}, and \ref{sun_bi}.  Because of its rich term structure, Fe contributes a 
significant fraction of the total line opacity, particularly in the UV band.  
Singly ionized Fe is the dominant stage, but in the LTE model
\ion{Fe}{2}/\ion{Fe}{1} is less than ten  
throughout the outer atmosphere.  Therefore, despite its minority status, \ion{Fe}{1}
contributes significant opacity, as a perusal of the line identifications in Figs. 
\ref{sun_atl1} to \ref{sun_atl3} shows.  As can be seen in
Fig. \ref{sun_bi}, the departure co-efficients, $b_{\rm i}$, 
including that for the ground state, are less 
than unity throughout much of
the atmosphere.  We emphasize again that these $b_{\rm i}$ values are with respect to 
$n_{\rm i}^*$ values that are computed with the {\it actual} value of $n_{\rm e}$ and ground
state population of the next higher ionization stage (see section \ref{s_modelcomp}).  
  Therefore, Fig. \ref{sun_bi} indicates that Fe is over-ionized with respect to the LTE
case.  
As a result, all the \ion{Fe}{1} lines are weakened with respect
to the LTE case such that the total opacity is reduced, particularly in the UV.  Therefore,
$F_{\lambda}$ is larger in the UV as compared to the LTE model.  This is a well-known effect that
has been found by previous independent investigations (see, for example, \citet{shchukina_t01}).

\subsubsection{Iron abundance and $b-f$ rates}

\begin{figure}
\plotone{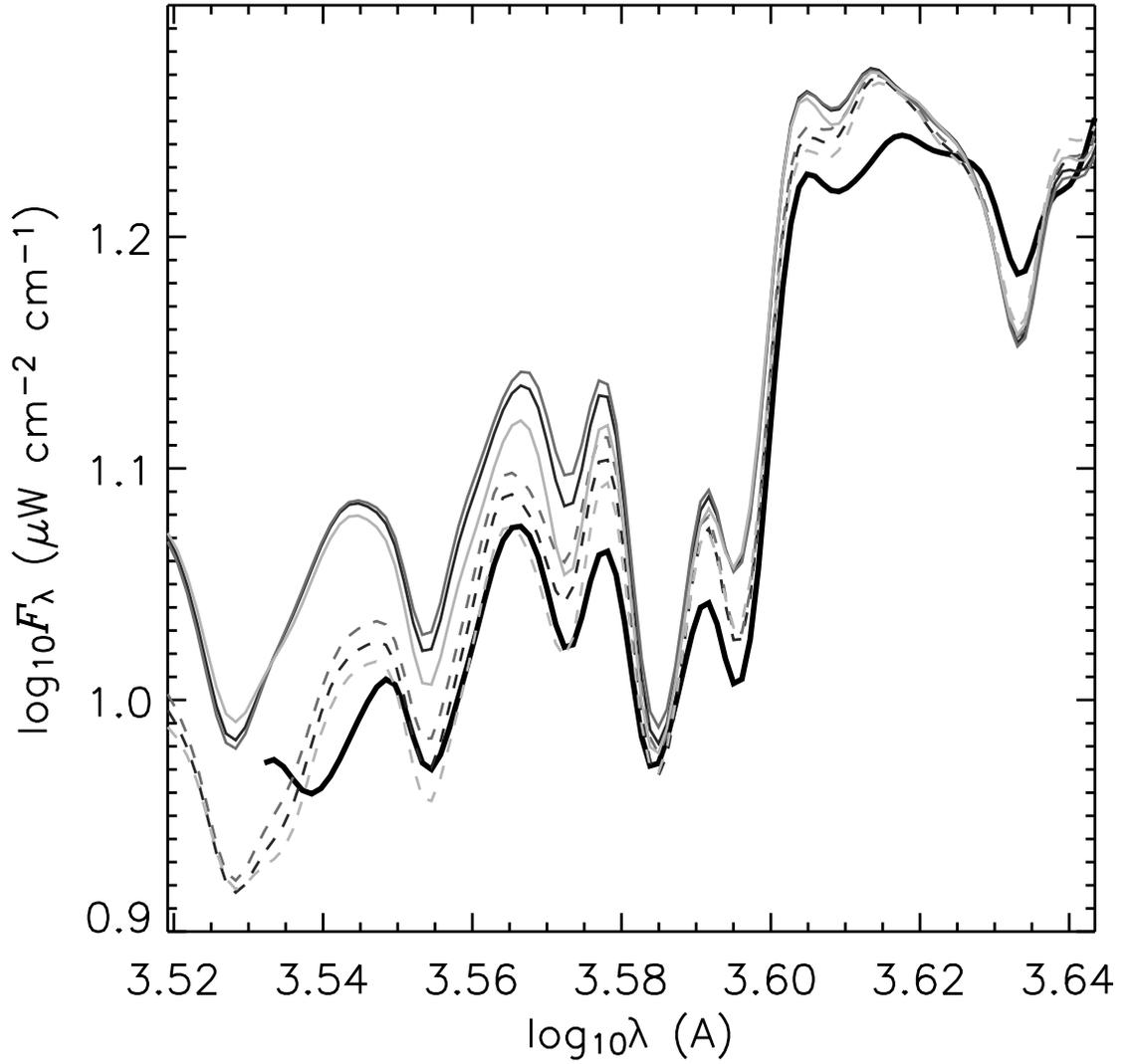}
\caption{As for Fig. \ref{sun_flxuv} except: theoretical distributions are shown for
LTE and NLTE$_{\rm Fe}$ models with $[{{\rm Fe}\over{\rm H}}]$ equal to 7.5 (the
originally adopted value) (dark line), 7.4 (medium line), and 7.7 (light line).
LTE models: dashed line, NLTE$_{\rm Fe}$ models: solid line. 
\label{fe_abund}}
\end{figure}

Veiling by many weak \ion{Fe}{1} lines plays a role in determining the UV band flux of late-type
stars, and the value of $[{{\rm Fe}\over{\rm H}}]$ for the Sun has been poorly constrained. 
To explore the effect of varying $[{{\rm Fe}\over{\rm H}}]$, we have re-calculated our models 
and their flux spectra with values 
set equal to the extrema of the range that has been recently published for the Sun, 
7.4 \citep{holweger_hk90} and 7.7 \citep{blackwell_bp84}.  In Fig. \ref{fe_abund} we show
the resulting UV band spectra.  The effect of varying $[{{\rm Fe}\over{\rm H}}]$
between the extrema is $\lambda$-dependent, being negligible at some
$\lambda$ values and as much as 0.3 dex at other values.  As expected, 
increased $[{{\rm Fe}\over{\rm H}}]$ leads to suppression of the UV flux
due to increased line blocking.  However, even with maximum $[{{\rm Fe}\over{\rm H}}]$
value the flux from the NLTE$_{\rm Fe}$ model is still significantly larger
than that observed.   

\paragraph{}

\begin{figure}
\plotone{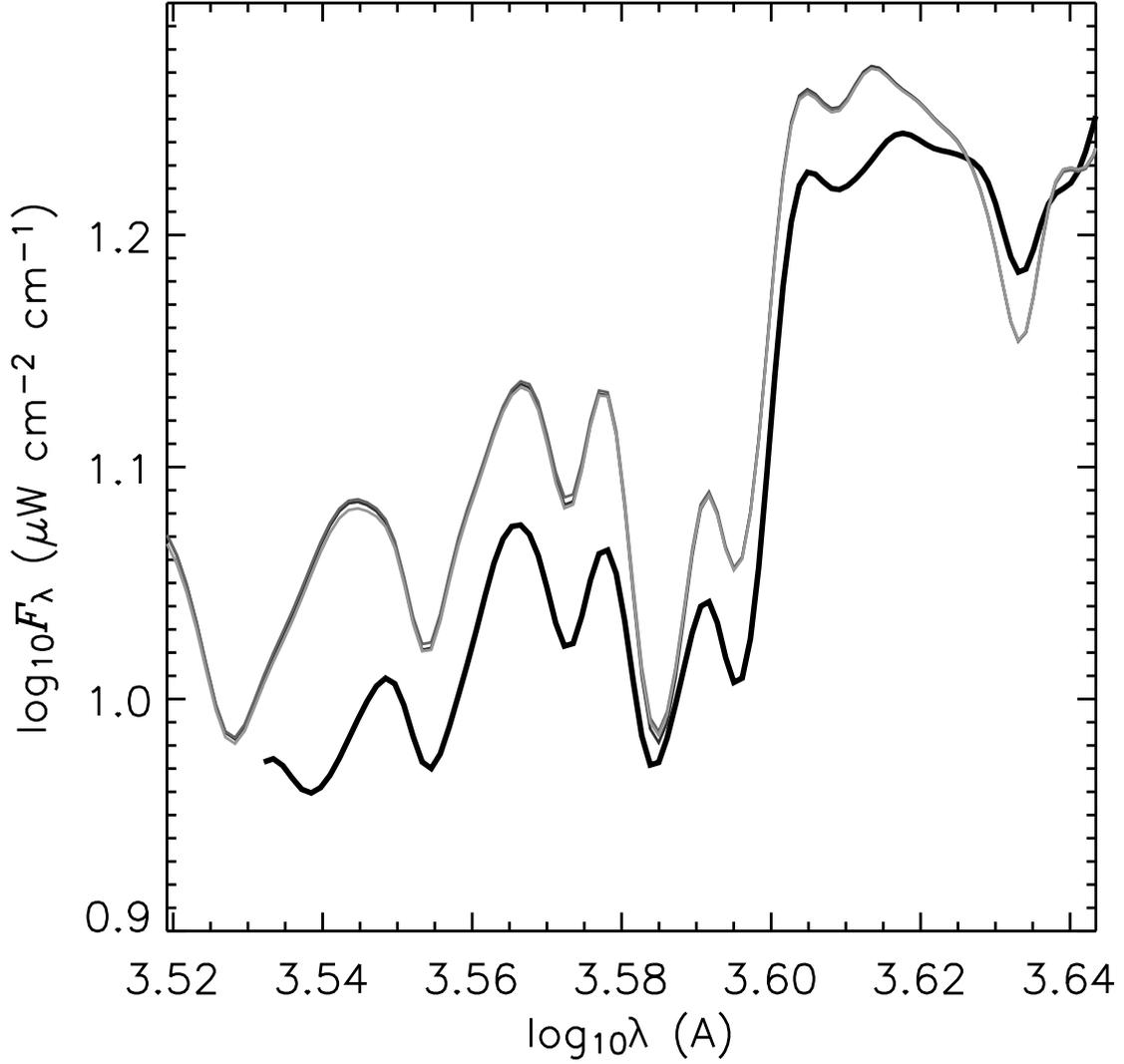}
\caption{As for Fig. \ref{sun_flxuv} except: theoretical distributions are shown for
NLTE$_{\rm Fe}$ models with $\sigma_{\rm b-f, Rad}$ and $\sigma_{\rm b-f, Col}$
for all \ion{Fe}{0}-group elements equal to Opacity Project or \citet{allen73} values, respectively
(dark line),
decreased by a factor of three and ten, respectively (medium line),
and increased by a factor of three and ten, respectively (light line).
\label{fe_bfcol}}
\end{figure}

\begin{figure}
\plotone{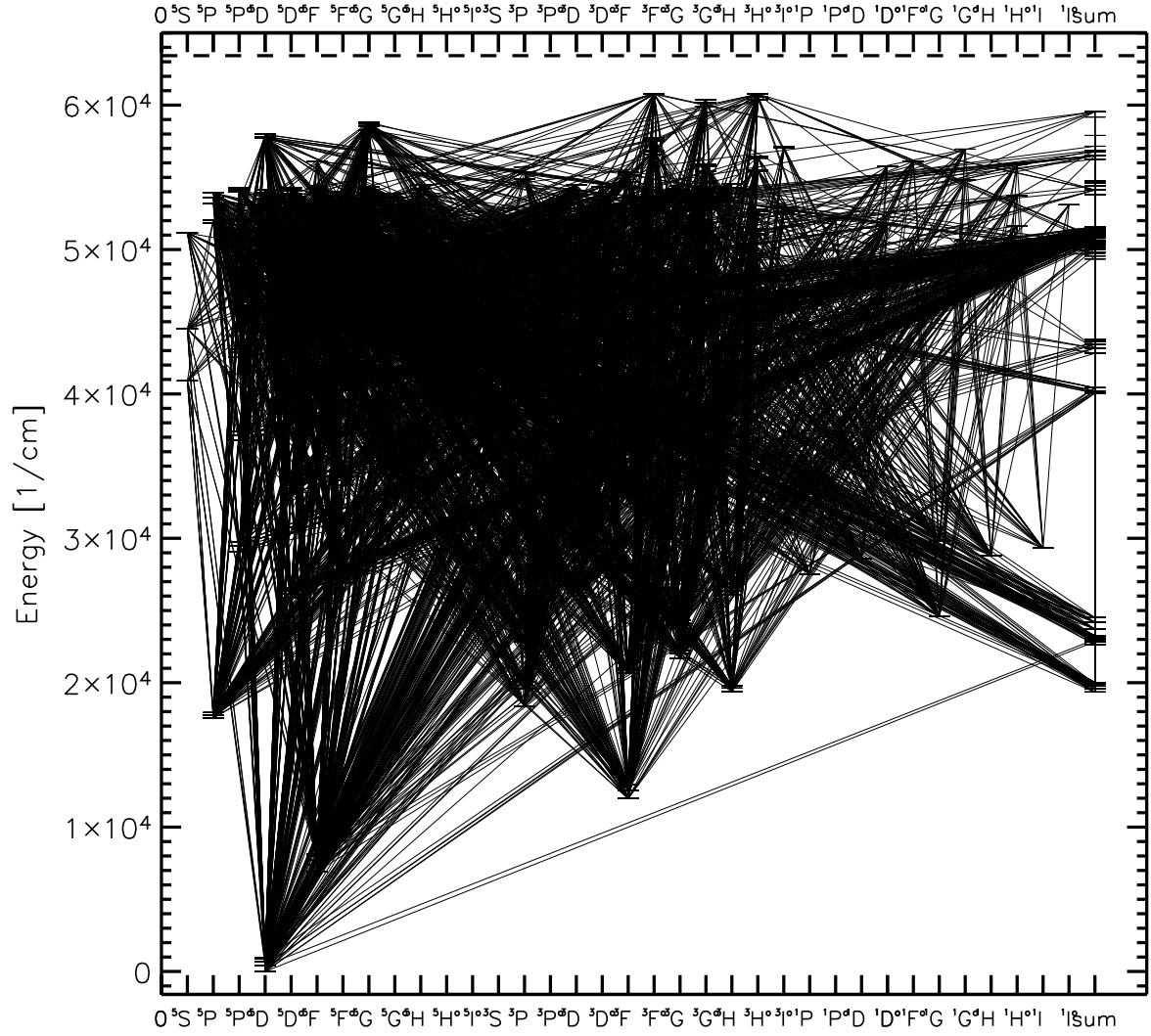}
\caption{Grotrian diagram of the model \ion{Fe}{1} atom used in 
our NLTE calculations.
\label{fe_grot}} 
\end{figure}

\begin{figure}
\plotone{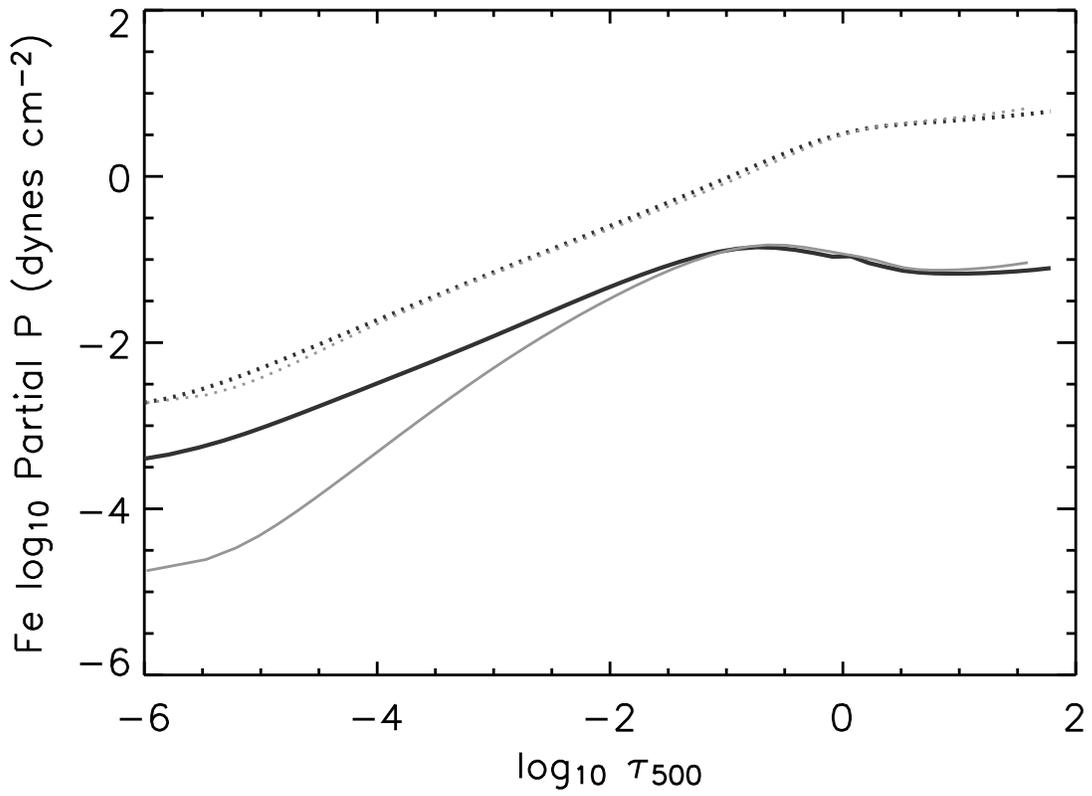}
\caption{Partial pressure of \ion{Fe}{1} (solid line) and \ion{}{2} (dotted line) in the
solar LTE (dark line) and the NLTE$_{\rm Fe}$ (light line) models.  
\label{sun_pp}} 
\end{figure}

\begin{figure}
\plotone{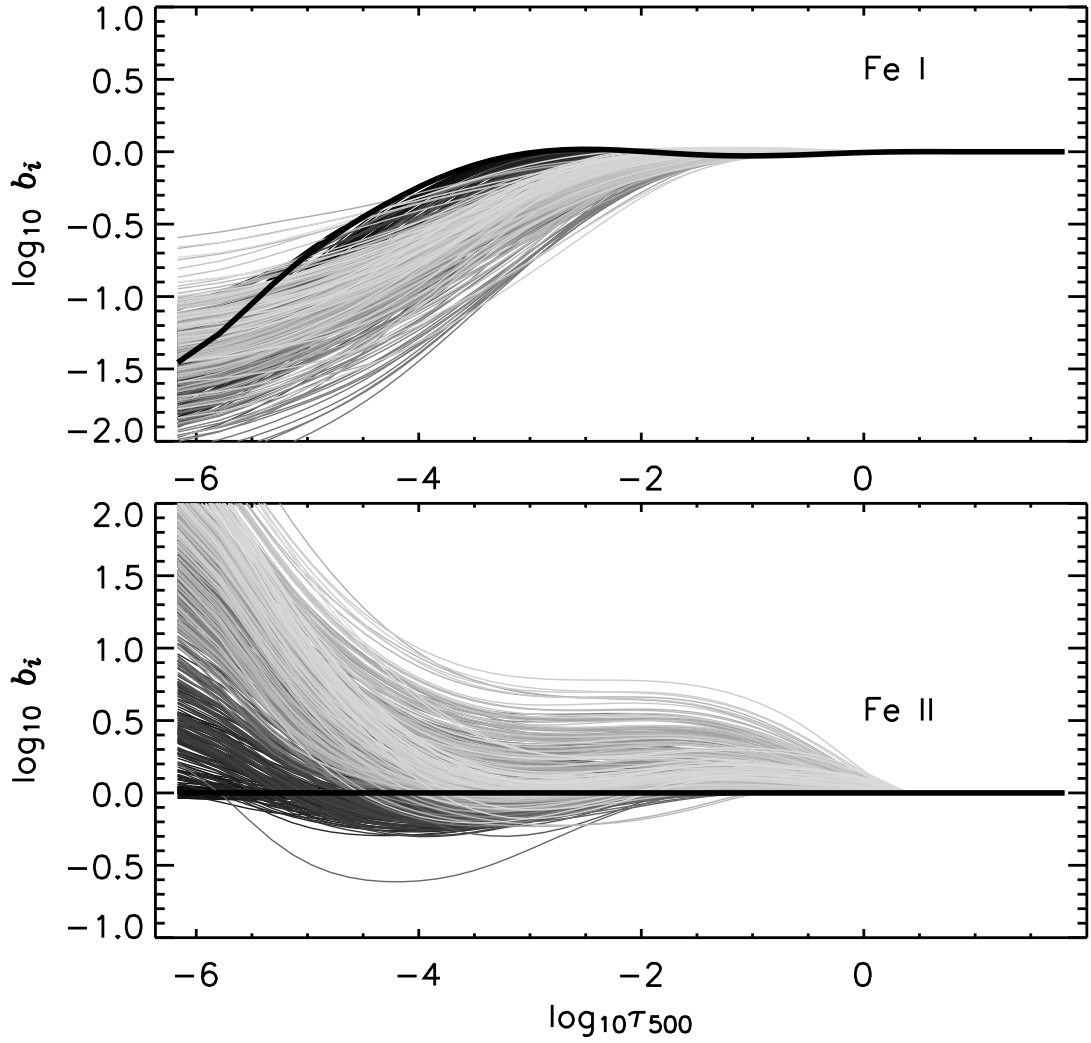}
\caption{NLTE departure coefficients for \ion{Fe}{1} and \ion{}{2} in the
solar NLTE$_{\rm Fe}$ model.  The 
ground state coefficient is shown with a thick black line.  The lighter
the color of the line the higher the energy, $E$, of the level.
\label{sun_bi}} 
\end{figure}

A complication that can compromise the value of NLTE modeling is that the solution is 
dependent on a larger variety of atomic data than is an LTE model, and these data 
are often uncertain by an order of magnitude or more.  In the case of
NLTE \ion{Fe}{1} over-ionization and the resulting UV flux level, the cross-section for both 
radiative and collisional ionization are important.  To investigate the dependence of 
our models on these rates we have calculated two variations on the NLTE$_{\rm Fe}$ model;
one in which the cross-sections for radiative and collisional $b-f$ processes,
$\sigma_{\rm b-f, Rad}$ and $\sigma_{\rm b-f, Col}$,
for {\it all} neutral stage \ion{Fe}{0}-group elements are {\it increased} by factors of
three and ten, respectively, and another in which they are {\it decreased} by a factors of
three and ten.  These perturbation factors adopted are the same
as those adopted by \citet{shchukina_t01}.
These two models represent the two conspiracies of error that would have 
maximum effect (eg. {\it all} rates for {\it all} neutral \ion{Fe}{0}-group species 
being maximally underestimated).  In Fig. \ref{fe_bfcol} we show the resulting
spectra in the UV band.  It can be seen that the perturbation affects the spectra negligibly, and cannot explain the discrepancy with the observed spectrum.

\subsection{Moderately high resolution spectrum \label{sect_hires}}

  Figs. \ref{sun_atl1} to \ref{sun_atl3} show the comparison between
the observed flux spectrum as measured by
\citet{kuru_fb} and medium resolution synthetic spectra computed with
our LTE and NLTE models for three 15 \AA~ bands 
throughout the UV and violet spectral region.  
The plots are annotated with the identity of the strongest line at each wavelength
where there is significant line opacity.

\paragraph{}

The synthetic spectra of all models were rectified by division by a pure continuum spectrum
that was synthesized with line opacity excluded with the corresponding model.  
We compare the observed and computed spectra at a resolution, $R$, of 200\,000 to emulate
the typical quality of the data for solar type stars.
Our purpose here is to assess the overall fit to typical observed stellar spectra
rather than to fit subtle details such as isotopic shifts or hyper-fine splitting. 
To ensure accurate registration for $\lambda$-wise differencing, the $\lambda$ scale of the
synthetic spectrum was empirically re-calibrated 
with a linear dispersion relation, the coefficients of which were
determined by minimizing the RMS deviation between our high resolution
$F_\lambda$ distribution as computed with the LTE model and the 
observed solar flux spectrum. 
We note that the Sun has a $v\sin i$ value of 2 km s$^{-1}$ that should
be taken into account in our synthetic spectra.  However, computational
constraints limit our NLTE spectrum synthesis to $R$ values that are too 
small for meaningful convolution with a kernel of 2 km s$^{\-1}$ width.  
However, we restrict our analysis to a discussion of spectral features
on a scale that is more gross than that of rotational broadening, and
to a strictly {\it differential} comparison of LTE and NLTE fits. 

\paragraph{}

An inspection of the identities of the lines 
in Fig. \ref{sun_atl1} through \ref{sun_atl3} reveals immediately
why the NLTE$_{\rm Fe}$ models differ from the LTE models much more than
the NLTE$_{\rm Light}$ models do.  Almost every line strong enough to meet
our criterion for being labeled, including almost all the lines with
extended damping wings, is an \ion{Fe}{1} line, and the remainder are largely
\ion{Fe}{0}-group lines.  Furthermore, there is an unlabeled veil of weak \ion{Fe}{1}
lines that collectively serves as a pseudo-continuous opacity in the UV.
This pervasiveness and dominance of Fe opacity in the violet and near UV bands
is a well known phenomenon.
Clearly, proper treatment of the NLTE \ion{Fe}{1} excitation equilibrium
and the \ion{Fe}{1}/\ion{Fe}{2} ionization equilibrium are of
dominant importance for accurate modeling of this region.  Of
secondary importance are other \ion{Fe}{0}-group elements such as Ti, Cr, and Ni.
We note that the first two ionization stages of Ti, Mn, Co, and Ni are included 
in the NLTE treatment
in our NLTE$_{\rm Fe}$ models.  

\paragraph{}

 Fig. \ref{sun_rms} shows the RMS deviation ($\sigma$) of the rectified
moderately high resolution model spectra
from the observed flux spectrum for the LTE and NLTE models as 
calculated for a 50\AA~ running mean throughout
the 3200 to 5000\AA~ range.  
It can be seen that the quality of the fit for all models
deteriorates significantly toward shorter $\lambda$ values.  The reason 
is that the density of spectral lines increases with decreasing $\lambda$.
As a result, the quality of the fit becomes increasingly sensitive to 
deficiencies in both the line list and the physical realism with
which line formation is modeled.  Also, as $\lambda$
decreases below the Wien peak of the black-body distribution for the 
Sun's $T_{\rm eff}$ (5000 \AA) the absolute $F_\lambda$ value becomes increasingly
sensitive to the temperature structure of the model atmosphere, and this
may indirectly effect the line formation.  Finally, although the pseudo-continuum does
not go completely into emission until $\lambda < 2500$ \AA, an increasing
number of strong lines may have chromospheric emission in their cores as $\lambda$
decreases.  

\paragraph{}

Fig. \ref{sun_rms2} contains similar plot, in which the steep global $\lambda$ dependence of the $\sigma(\lambda)$
surface has been removed by $\lambda$-wise
subtraction of the $\sigma(\lambda)$ values of the LTE model, 
$\sigma(\lambda)-\sigma(\lambda, LTE)$.  
Fig. \ref{sun_rms2} allows a direct assessment of how well
the fit of each model compares to that of the LTE model as a function
of $\lambda$.
Note that, to make the entire $\sigma$ surface
visible, the $\lambda$ axis has been reversed with respect to that of 
Fig. \ref{sun_rms}.  
It can be seen that $\sigma-\sigma(LTE)$ for the NLTE synthetic spectra 
are scattered within $\pm\approx 0.02$ rectified continuum units of 
that of the LTE model.  This indicates that the NLTE spectra provide
a better fit to the detailed shape of the moderately high resolution
observed spectrum 
at some $\lambda$ intervals, and a worse one at others.  The NLTE$_{\rm Fe}$ 
spectrum shows a larger range of  $\sigma-\sigma(LTE)$ than does the 
NLTE$_{\rm Light}$ spectrum, which indicates that inclusion of \ion{Fe}{0}-group 
elements in the NLTE SE exaggerates further still the goodness
of fit at some $\lambda$ intervals, and the badness of the fit at others.
However, given that $\sigma-\sigma(LTE) < 0.02$ at most $\lambda$ points,
we may conclude that both NLTE synthetic spectra give a fit to the detailed
shape of the observed spectrum that does not differ significantly from  
that of the LTE spectrum.  We conclude that, within the limits of our modeling,
globally fitting the detailed shape
of the spectrum is not as good a test of NLTE models as fitting the
overall flux level.

\subsubsection{van der Waals broadening}

\begin{figure}
\plotone{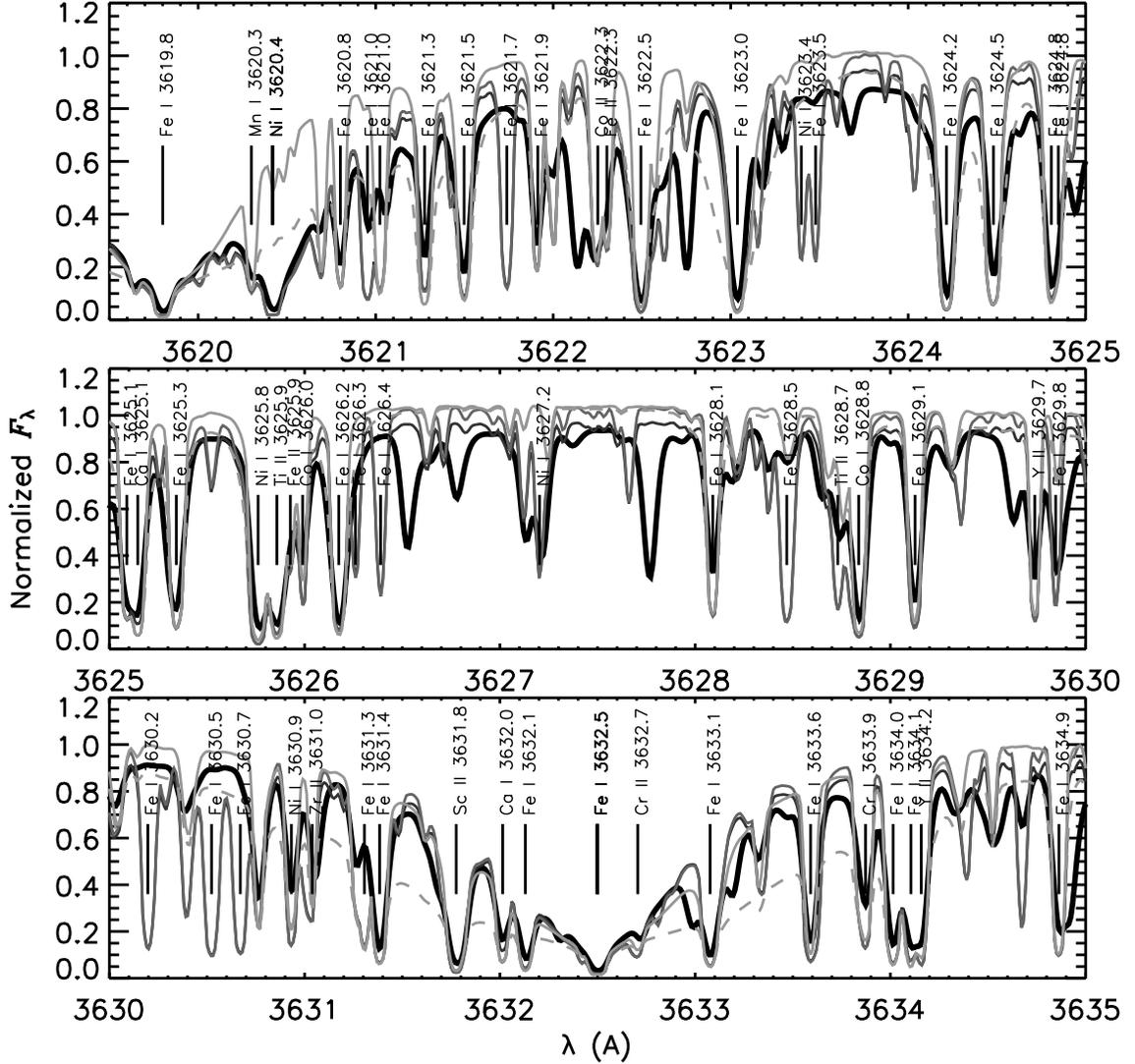}
\caption{Comparison of observed solar flux \citep{kuru_fb}
(thick black line) in the 3620 to 3635 \AA~ region (just below the Balmer jump) 
and synthetic spectra computed with the LTE model
(thin dark line), NLTE$_{\rm Light}$ (medium line) and NLTE$_{\rm Fe}$ (light line) models with no $\gamma_6$ enhancement, and with the NLTE$_{\rm Fe}$ model
with a $\gamma_6$ enhancement factor of 1.8 (dashed line). 
\label{sun_atl1}}
\end{figure}

\begin{figure}
\plotone{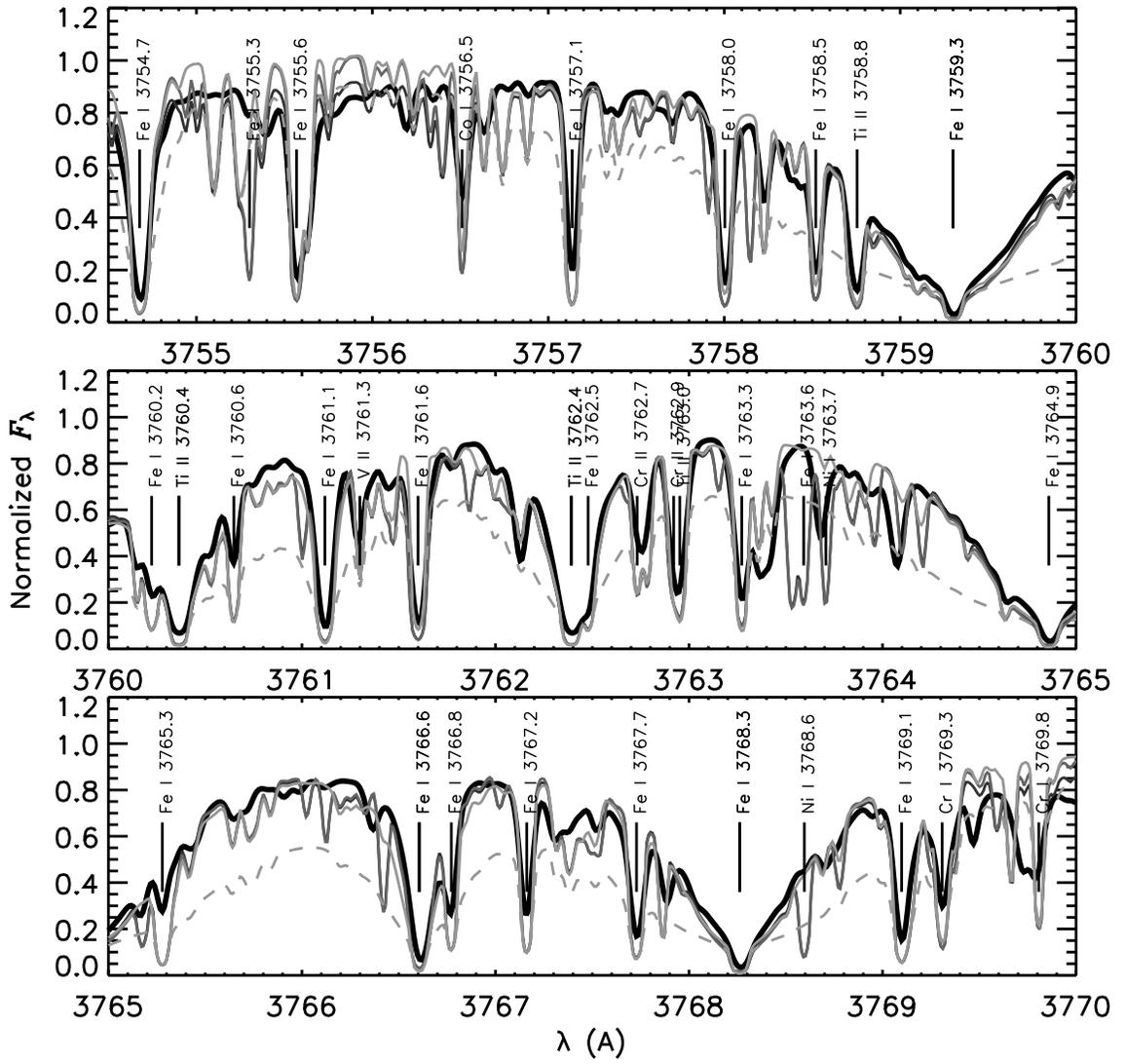}
\caption{Same as Fig. \ref{sun_atl1}, but for 3755 to 3770\AA~ region.  
\label{sun_atl2}}  
\end{figure}

\begin{figure}
\plotone{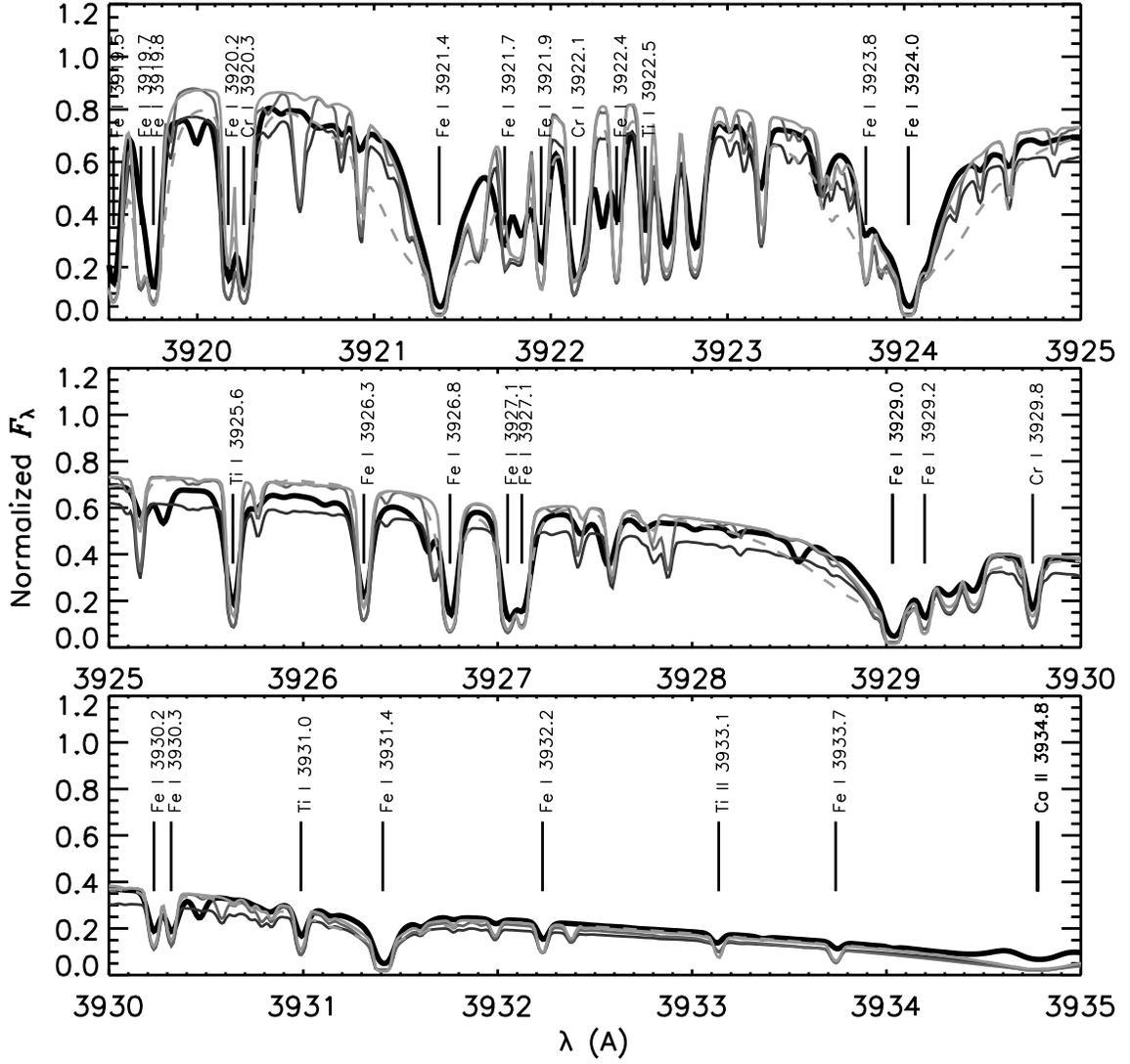}
\caption{Same as Fig. \ref{sun_atl1}, but for 3920 to 3935\AA~ region.  The 
broad absorption line that spans the lowest two panels is the \ion{Ca}{2} $K$
line. \label{sun_atl3}}  
\end{figure}

\begin{figure}
\plotone{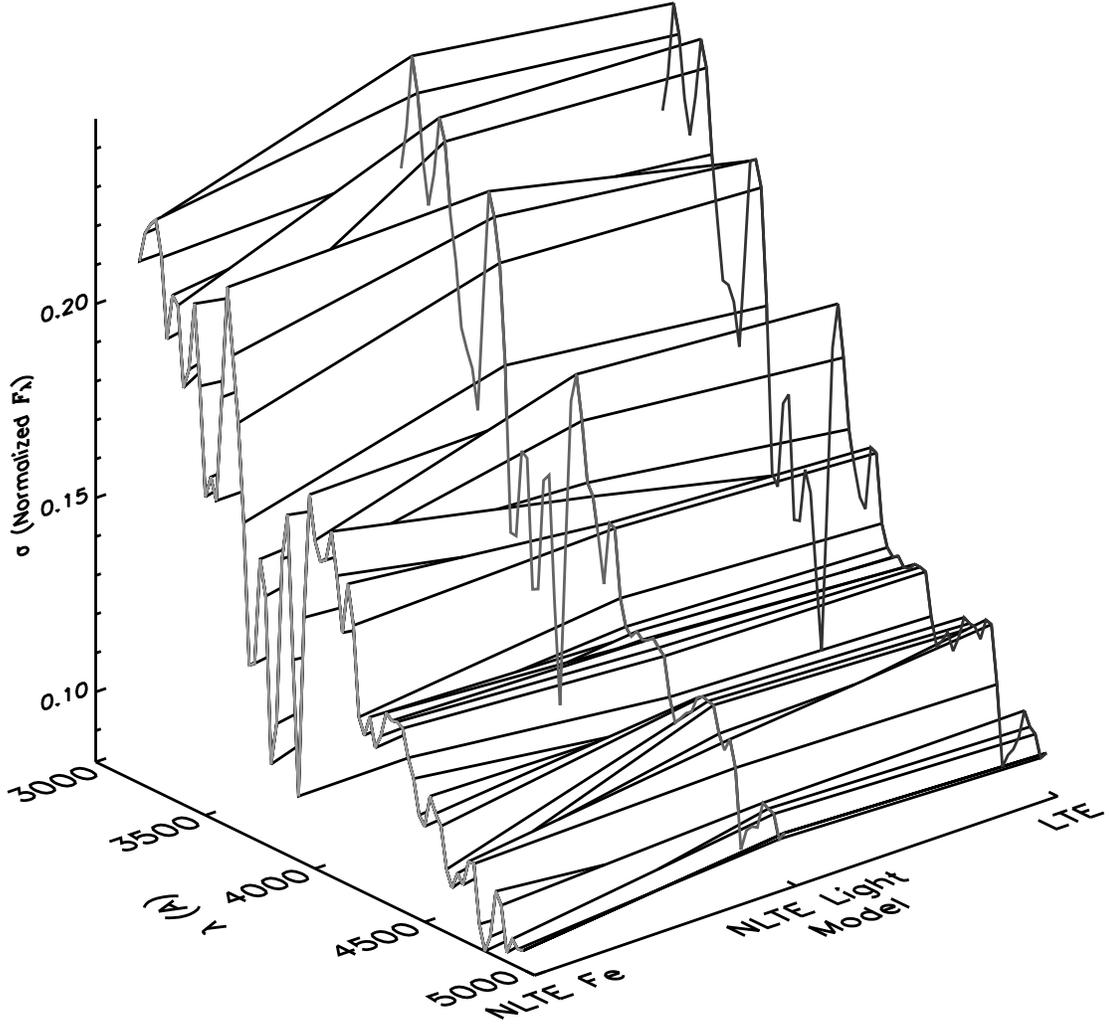}
\caption{RMS deviation ($\sigma$) of the computed high
resolution rectified flux distribution ($F_\lambda(\lambda)$) from the 
high resolution spectrum measured by \citet{kuru_fb} in the UV 
and blue bands,
as calculated for a running 50\AA~ window.  Deviations are shown for
the LTE (dark line), NLTE$_{\rm Light}$ (medium line), and NLTE$_{\rm Fe}$
(light line) models.  \label{sun_rms}}
\end{figure}

\begin{figure}
\plotone{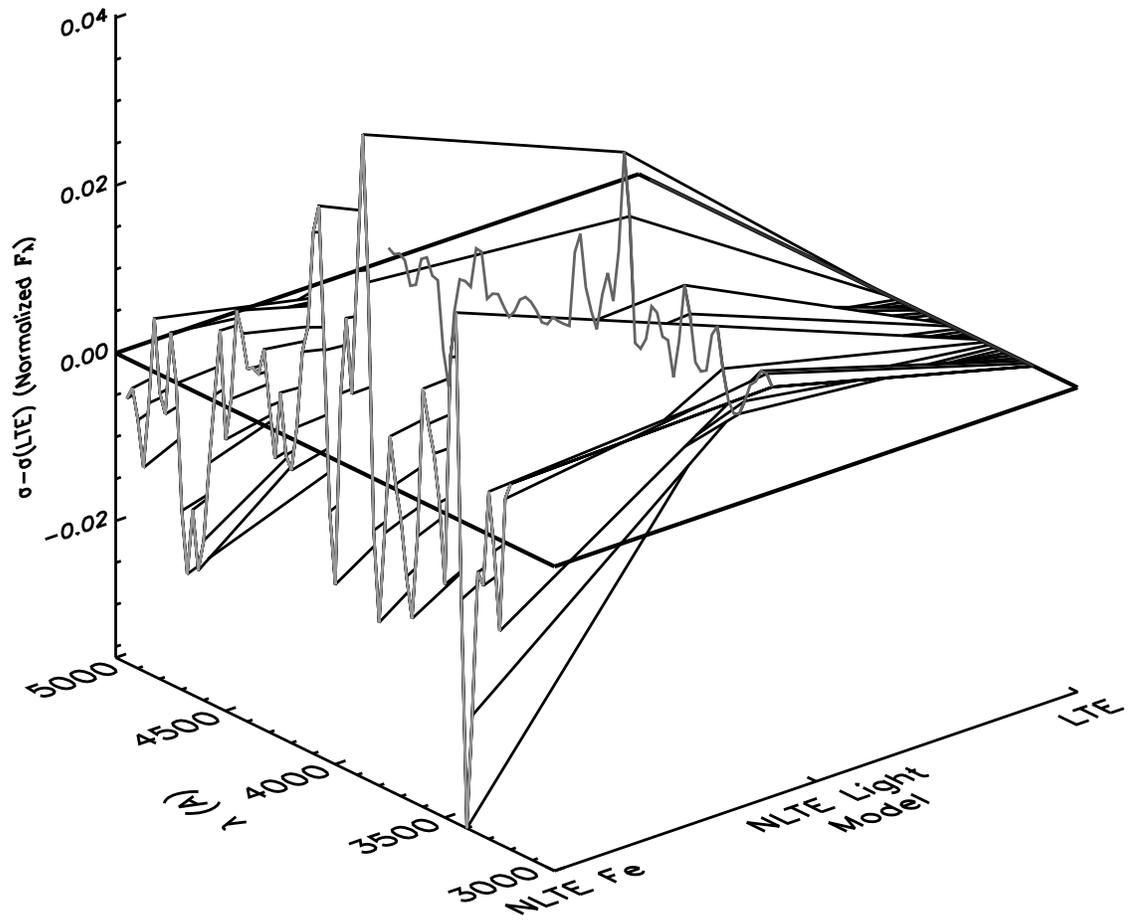}
\caption{As for Fig. \ref{sun_rms}, but with the RMS deviation ($\sigma$)
normalized by $\lambda$-wise subtraction of $\sigma$ for the LTE model.
Note that, to allow the clearest view of the $\sigma$ surfaces, the 
wavelength ($\lambda$) axis in this figure is reversed with respect to that of 
Fig. \ref{sun_rms}.
\label{sun_rms2}}
\end{figure}

Voigt profiles were calculated for the spectral lines that take into account 
van der Waals
(VW) broadening due to collisions with \ion{H}{1}, which dominates the
line broadening in the solar atmosphere.  The VW broadening
parameter, $\gamma_6$, is calculated with the Unsoeld
approximation \citep{unsoeld}.  To fit the profiles of damped lines, many previous
investigators find it necessary to 
enhance the value of $\gamma_6$, {\it ad hoc}, by an arbitrary factor, $E_6$,
that usually ranges from 1 to 3 (see \citet{kostik_sr96} for a 
detailed discussion).  The physical meaning of the enhancement is unclear,
and one possibility is that, like micro-turbulence ($\xi_{\rm T}$), 
it is at least partially a ``fudge'' factor to account for inhomogeneities
in the solar (or stellar) atmosphere \citep{shchukina_t01}.     
We have 
computed our spectra with an $E_6$ value of unity; ie. {\it no} enhancement.
We also show in Figs. \ref{sun_atl1} through \ref{sun_atl3}
a spectrum computed with the NLTE$_{\rm Fe}$ model and a more traditional $E_6$
value of 1.8, which provides a closer match to the observed $F_\lambda$ level (see 
section \ref{sect_absflux}.  
It can be seen that in the near UV band,
 the traditional enhancement gives
computed line profiles that are systematically too broad, whereas, an
enhancement factor of unity provides a close match to all the damped lines.   
The only exception is the very broad \ion{Ca}{2} $\lambda 3934.8$ line 
(Fig. \ref{sun_atl3}), which is equally well fit with both values of 
$E_{\rm 6}$.  

\paragraph{}

For any other star, one might argue that the necessity of neglecting
the $\gamma_6$ enhancement is masking an inadequacy in the model
that exaggerates the pressure broadening, such as an adopted value of 
$\log g$ or $[A/H]$ that is too large.  However, in the case of the Sun
the atmospheric parameters are known to very high precision.
It is noteworthy that most of 
the damped lines in Figs. \ref{sun_atl1} and \ref{sun_atl2} are \ion{Fe}{1}
lines.  
\citet{blackwell_ls95} found that the value of $E_6$ that provides
the best fit to \ion{Fe}{1} lines in the visible band depends on 
multiplet number, with lines arising from multiplet numbers less that 50
having best fit $E_6$ values less than 1.1.  In the 3000 to 4000 \AA~
band, 26 out of about 140 \ion{Fe}{1} lines of $\log gf$ value greater than -1
listed in the NIST multiplet table arise from multiplet numbers less than 50. 
Similarly, \citet{bensby_fl03} investigated the variation of derived abundances
for F and G dwarfs with input atomic parameters and adopted a relatively low
value of $E_6=1.4$ for \ion{Fe}{1} lines of lower excitation potential
greater than 2.6 eV.
Another possibility is that the systematic 
over-damping of these lines in the LTE spectrum
synthesis with a $\gamma_6$ enhancement of 1.8 is due to neglect of the 
NLTE over-ionization of Fe.  However,
comparison of the LTE and NLTE$_{\rm Fe}$ synthetic spectra in Figs. \ref{sun_atl1} 
and \ref{sun_atl2} shows that,
while NLTE over-ionization may lift the veil of weak \ion{Fe}{1} lines that
collectively block UV flux, it is not enough to significantly weaken 
the profiles of damped \ion{Fe}{1} lines. 


\section{Conclusions \label{sec_con}}

   We have presented atmospheric models and synthetic flux spectra
for a late-type MS star with the parameters of the Sun, that represent three 
degrees of realism in the treatment of the equilibrium state of the gas and 
the radiation field.  We have studied separately NLTE effects in \ion{Fe}{0}-group
elements and light metals on the model
structure and emergent flux, and the ability of these models to 
fit both the broad-band absolute $F_\lambda(\lambda)$ level and 
moderately high resolution spectra.

\paragraph{}

The theoretical $T(\tau)$ structure of our most realistic model (NLTE$_{\rm Fe}$)
agrees to within 100 K with both the more approximate theoretical NLTE line blanketed model of
\citet{anderson_89} ({\sc PAM}) and the {\it LTE} semi-empirical structure found by of 
\citet{holmul} (HOLMUL).  The LTE and NLTE$_{\rm Light}$ models are in closer agreement
to the theoretical LTE model of \citet{kurucz92a} and the {\it LTE} semi-empirical structure found by
\citet{grev_s99} ({\sc HOLMUL'}).  The agreement between the NLTE$_{\rm Fe}$ model and
the {\sc PAM} model is remarkable vindication of the approximate {\it multifrequency/multigray}
method developed by \citet{anderson_89}.  The comparison with the two LTE semi-empirical models is of
particular interest: {\sc HOLMUL'} is based on an LTE analysis of the \ion{Fe}{1} excitation
equilibrium, whereas {\sc HOLMUL} is based on a wider variety of spectral features and center-to-limb
variation.  Therefore, the former may fall prey to the ``self-fulfilling prophecy'' effect 
described by \citet{rutten86}: adoption of LTE in the treatment of \ion{Fe}{0} leads to an
inferred $T_{\rm kin}$ structure that produces the observed spectrum when \ion{Fe}{0} is treated in LTE. 

\paragraph{}
 
The NLTE effects of \ion{Fe}{0}-group 
elements on the model structure and $F_\lambda$ distribution are much more 
important than the NLTE effects of all the light metals combined,
and serve to substantially increases the violet and near UV
$F_\lambda$ level as a result of NLTE Fe over-ionization.
Based on calculations of LTE and NLTE $F_\lambda$ distributions 
with the semi-empirical $T_{\rm kin}(\tau)$ structure of \citet{grev_s99}, 
and on calculation of the LTE $F_\lambda$ distribution with the NLTE$_{\rm Fe}$
$T_{\rm kin}(\tau)$ structure, the discrepancy between the observed and predicted
$F_\lambda$ level is much too large to be due to errors in the model structure. 
These results suggest that there may {\it still} be
important UV opacity missing from the models.  
We know that nature is not obliged to be in LTE, and, therefore, on general
principle, it should be modeled in NLTE.  If NLTE calculations
sometimes {\it worsen} the fit of models to observational data, that 
may serve
to expose other modeling inadequacies that had been partially hidden by LTE 
modeling.
We also find that 
the RMS deviation, $\sigma(\lambda)$, of the moderately high resolution 
synthetic 
spectrum from the observed spectrum is changed randomly by $\pm 0.02$ 
rectified continuum units by the adoption of NLTE.  We conclude from the
latter that within the limits of the present models, the statistical 
goodness of fit to many line profiles is not a good discriminator of merit
between LTE and NLTE models. 

\paragraph{}

The quality of the fit to the UV $F_\lambda$ level is sensitive to
the extent to which strong lines are damped by van der Waals broadening,
and the notorious vdW enhancement parameter, $E_6$, is species (and transition) 
dependent.  A traditional value of $E_6$ of 1.8 gives a close fit to the 
observed $F_\lambda$ distribution but an unacceptably poor fit to the
detailed shape of damped lines.  On the other hand, a value of unity
({\it ie.} no enhancement) gives a close fit to the many damped \ion{Fe}{1}
lines in the near UV while leading to a predicted $F_\lambda$ distribution
that is too bright.



\acknowledgments

CIS gratefully acknowledges funding from the Natural Sciences and Engineering 
Research Council of Canada (grant no. 264515-03), Saint Mary's University, 
and from the Charles E. Schmidt College of Science at Florida Atlantic 
University.
This work was supported in part by NSF grants AST-9720704 and  
AST-0086246, NASA
grants NAG5-8425, NAG5-9222, as well as NASA/JPL grant 961582 to the  
University
of Georgia.  This work was supported in part by the P\^ole Scientifique  
de
Mod\'elisation Num\'erique at ENS-Lyon.  Some of the calculations  
presented in
this paper were performed on the IBM pSeries 690 of the Norddeutscher  
Verbund
f\"ur Hoch- und H\"ochstleistungsrechnen (HLRN), on the IBM SP  
``seaborg'' of
the NERSC, with support from the DoE, and on the IBM SP ``Blue  
Horizon'' of the
San Diego Supercomputer Center (SDSC), with support from the National  
Science
Foundation.  We thank all these institutions for a generous allocation  
of
computer time.





\clearpage

\clearpage

\clearpage 

\clearpage 

\clearpage

\clearpage

\clearpage 

\clearpage 

\clearpage 

\clearpage 

\clearpage 

\clearpage 

%
%
%
%
\clearpage 

\clearpage 

\clearpage 

\begin{deluxetable}{lrrrrrrr}
\footnotesize
\tablecaption{Species treated in Non-Local Thermodynamic Equilibrium (NLTE)
in the NLTE$_{\rm Light}$ and NLTE$_{\rm Fe}$ models.
Each ionization stage is labeled with the number of energy levels and bound-bound
transitions included in the statistical equilibrium rate equations.  Note that
this table shows only a sub-set of the total number of species that are 
currently treatable in statistical equilibrium by {\sc PHOENIX. }
}
\label{t1d}
\tablecolumns{8}
\tablewidth{0pt}
\tablehead{
\colhead{Element} & Model & \multicolumn{3}{c}{Ionization Stage} \\
\colhead{} & \colhead{} & \colhead{\ion{}{1}} & \colhead{\ion{}{2}} & \colhead{\ion{}{3}} }
\startdata
H   & NLTE$_{\rm Light}$, NLTE$_{\rm Fe}$     &  80/3160 &\nodata &\nodata  \\
He  & NLTE$_{\rm Light}$, NLTE$_{\rm Fe}$     &  19/37 & \nodata &\nodata  \\
Li  & NLTE$_{\rm Light}$, NLTE$_{\rm Fe}$     &  57/333 & 55/124 &\nodata  \\
C   & NLTE$_{\rm Light}$, NLTE$_{\rm Fe}$     &  228/1387 & \nodata & \nodata \\
N   & NLTE$_{\rm Light}$, NLTE$_{\rm Fe}$     &  252/2313 & \nodata & \nodata \\
O   & NLTE$_{\rm Light}$, NLTE$_{\rm Fe}$     &  36/66 & \nodata & \nodata \\
Ne  & NLTE$_{\rm Light}$, NLTE$_{\rm Fe}$     &  26/37 &\nodata &\nodata \\
Na  & NLTE$_{\rm Light}$, NLTE$_{\rm Fe}$     &  53/142 & 35/171 & \nodata \\
Mg  & NLTE$_{\rm Light}$, NLTE$_{\rm Fe}$     &  273/835 & 72/340 & \nodata \\
Al  & NLTE$_{\rm Light}$, NLTE$_{\rm Fe}$     &  111/250 & 188/1674 & \nodata \\
Si  & NLTE$_{\rm Light}$, NLTE$_{\rm Fe}$     &  329/1871 & 93/436 & \nodata \\
P   & NLTE$_{\rm Light}$, NLTE$_{\rm Fe}$     &  229/903 & 89/760 & \nodata \\
S   & NLTE$_{\rm Light}$, NLTE$_{\rm Fe}$     &  146/439 & 84/444 & \nodata \\
K   & NLTE$_{\rm Light}$, NLTE$_{\rm Fe}$     &  73/210 &  22/66 & \nodata\\
Ca  & NLTE$_{\rm Light}$, NLTE$_{\rm Fe}$     &  194/1029 & 87/455 & 150/1661 \\
Ti  & NLTE$_{\rm Fe}$     &  395/5279 & 204/2399 & \nodata \\
Mn  & NLTE$_{\rm Fe}$     &  316/3096 & 546/7767 & \nodata \\
Fe  & NLTE$_{\rm Fe}$     &  494/6903 & 617/13675 & \nodata\\
Co  & NLTE$_{\rm Fe}$     &  316/4428 & 255/2725 & \nodata\\
Ni  & NLTE$_{\rm Fe}$     &  153/1690 & 429/7445 & \nodata\\
\enddata
\end{deluxetable}

\begin{table}
\caption{Levels of modeling realism.}
\label{t2}
\begin{tabular}{lll}
\tableline
Degree of NLTE          & Model designation\\
\tableline
None                    & {LTE} \\
Light metals only       & {NLTE$_{\rm Light}$} \\
Light metals \& \ion{Fe}{0}-group   & {NLTE$_{\rm Fe}$} \\
\tableline
\end{tabular}
\end{table}

\begin{table}
\caption{Iron abundance and secondary stellar parameters of {\sc PHOENIX} models that are compared to 
models of other authors.}
\label{t3}
\begin{tabular}{llll}
\tableline
Author/Model                 & $[{{\rm Fe}\over{\rm H}}]$ & $l/H_{\rm P}$ & $\xi_{\rm T}$ (km s$^{-1}$) \\ 
\citet{kurucz92a}/{\sc ASUN}  & 7.63                       & 1.25          & 1.5             \\
\citet{anderson_89}/{\sc PAM}  & 7.50                       & 0.0           & 1.5             \\
\tableline
\end{tabular}
\end{table}

\end{document}